\documentclass[10pt, conference]{IEEEtran}
\IEEEoverridecommandlockouts
\usepackage{cite}
\usepackage{amsmath,amssymb,amsfonts}
\usepackage{algorithmic}
\usepackage{graphicx}
\usepackage{textcomp}
\usepackage{xcolor}

\usepackage[hidelinks]{hyperref}
\usepackage{bbm}
\usepackage{array,multirow}
\usepackage{makecell}
\usepackage{siunitx}
\usepackage{booktabs}
\usepackage{arydshln}
\usepackage[capitalise]{cleveref}
\newcommand{\eg}{e.g., }
\newcommand{\ie}{i.e., }

\newrobustcmd{\B}{\bfseries}

\def\BibTeX{{\rm B\kern-.05em{\sc i\kern-.025em b}\kern-.08em
    T\kern-.1667em\lower.7ex\hbox{E}\kern-.125emX}}
\begin{document}

\title{Joint semi-supervised and contrastive learning enables domain generalization and multi-domain segmentation}

\author{
\IEEEauthorblockN{
Alvaro Gomariz$^{1 *}$,
Yusuke Kikuchi$^{2}$,
Yun~Yvonna Li$^{1}$,
Thomas Albrecht$^{1}$,
Andreas Maunz$^{1}$,
Daniela Ferrara$^{2}$,\\
Huanxiang Lu$^{1}$,
Orcun Goksel$^{3}$
}
\IEEEauthorblockA{
\\
$^1$F Hoffmann-La Roche AG, Basel, Switzerland \\
$^2$Genentech Inc, California, United States \\
$^3$Department of Information Technology, Uppsala University, Uppsala, Sweden
}
}
\maketitle

\begingroup\renewcommand\thefootnote{*}
\footnotetext{Corresponding author: \hyperlink{alvaro.gomariz@roche.com}{alvaro.gomariz@roche.com}}

\begin{abstract}
Despite their effectiveness, current deep learning models face challenges with images coming from different domains with varying appearance and content. 
We introduce SegCLR, a versatile framework designed to segment images across different domains, employing supervised and contrastive learning simultaneously to effectively learn from both labeled and unlabeled data. 
We demonstrate the superior performance of SegCLR through a comprehensive evaluation involving three diverse clinical datasets of 3D retinal Optical Coherence Tomography (OCT) images, for the slice-wise segmentation of fluids with various network configurations and verification across 10 different network initializations. 
In an unsupervised domain adaptation context, SegCLR achieves results on par with a supervised upper-bound model trained on the intended target domain. 
Notably, we discover that the segmentation performance of SegCLR framework is marginally impacted by the abundance of unlabeled data from the target domain, thereby we also propose an effective domain generalization extension of SegCLR, known also as zero-shot domain adaptation, which eliminates the need for any target domain information.
This shows that our proposed addition of contrastive loss in standard supervised training for segmentation leads to superior models, inherently more generalizable to both in- and out-of-domain test data. 
We additionally propose a pragmatic solution for SegCLR deployment in realistic scenarios with multiple domains containing labeled data. 
Accordingly, our framework pushes the boundaries of deep-learning based segmentation in multi-domain applications, regardless of data availability —-- labeled, unlabeled, or nonexistent.
\end{abstract}


\section{Introduction}

Routine assessment of eye conditions in clinics is being transformed by the applications of deep learning in ophthalmology. 
These methods facilitate quantitative analysis of 3D images of eyes acquired with  Optical Coherence Tomography (OCT) devices~\cite{fujimoto2016development}, which create volumetric datasets by stacking 2D slices known as B-scans.
The use of supervised learning in segmentation of retinal fluids, mainly via UNet~\cite{unet} based neural networks, has led to advances in diagnosis, prognosis, and a deeper understanding of eye diseases, such as neovascular age-related macular degeneration (nAMD) and Diabetic Macular Edema (DME)~\cite{schmidt2016paradigm,sahni2019machine,unetretina18,bogunovic2019retouch}. 
However, training such supervised deep neural networks requires large amounts of labeled data, the procurement of which is costly and sometimes infeasible.
Most importantly, such labeling needs to be repeated for each problem setting/domain; since
trained models often fail when inference data differs from labeled examples, so-called \emph{domain-shift}, \eg between images from different eye diseases and different OCT devices~\cite{schlegl2018fully}. 

Domain adaptation techniques aim to maintain a similar performance on images from new domains while reducing the annotation effort required in such newer datasets. 
Unsupervised domain adaptation aims to leverage information learned from a labeled data domain for applications in other domains where only unlabeled data is available. 
To this end, many deep learning methods have been proposed~\cite{wang2018deep}, mostly using generative adversarial networks, \eg to translate visual appearance across OCT devices~\cite{ren2021segmentation}.
Ultimately, it is desirable for learned models to implicitly generalize to other domains without requiring any new data, \ie ideally not even unlabeled data from a new domain.
This scenario, known as domain generalization~\cite{zhou2022domain} or zero-shot domain adaptation~\cite{peng2018zero}, provides huge advantages in real-world deployment by obviating any retraining or customization being required per test domain encountered. 

Advances in self-supervised learning have been increasingly successful in extracting informative features from images without accessing their labels, hence enabling the exploitation of much larger unlabeled datasets~\cite{caron2021emerging,simclr,chen2020big,chen2021exploring,grill2020bootstrap,he2020momentum,khosla2020supervised,oord2018representation}.
These methods usually employ different versions of Siamese networks~\cite{bromley1993signature} that can be applied to different inputs in parallel while sharing their weights. 
For self-supervision, typically different augmentations of an image are treated as positive samples, which are desired to map close-by or to a single point on a learned manifold, and therefore minimizing the feature distances between these positive samples when passed through a Siamese network. 
To prevent such optimization from collapsing to a constant output, different approaches have been proposed: 
For instance, \mbox{BYOL}~\cite{grill2020bootstrap} and \mbox{SimSiam}~\cite{chen2021exploring} employ a stop-gradient operation for one of their two encoders, with their main difference being how the other encoder is updated: BYOL uses the moving average of the parameters of the stop-gradient encoder, whereas SimSiam uses a Siamese architecture between its two encoders.
Another successful strategy, contrastive learning (CL), introduces the concept of negative pairs, the distances of which should meanwhile be maximized. 
SimCLR~\cite{simclr} is one of the most widely-adopted CL frameworks. 
Self-supervised learning is commonly used for model pretraining, typically based on natural images such as ImageNet~\cite{deng2009imagenet}.
Such pretrained model is subsequently fine-tuned or distilled for downstream tasks such as classification, detection, or segmentation~\cite{chen2020big}.
Notably, while BYOL is a more recent approach, when handling multiple domains, better results have been reported~\cite{zhang2022towards} with SimCLR, on which we build our method in this work.

Models pretrained with natural images are of limited use for medical applications, which often involve images with volumetric content and with substantially differing appearances from natural images.
This has led to recent application-specific approaches for CL pair generation in medical context~\cite{chaitanya2020contrastive,chen2021uscl}.
USCL~\cite{chen2021uscl} minimizes the feature distance between frames of the same ultrasound video, while maximizing the distance between frames of different videos, in order to produce pretrained models for ultrasound applications. 
USCL also proposes a joint semi-supervised approach, which simultaneously minimizes a contrastive and supervised \emph{classification} loss.
However, to be applicable for image segmentation, this method relies on subsequent fine-tuning, which is potentially sub-optimal for preserving the unlabeled information for the intended task of segmentation.
In fact, there exist little work on CL methods on image segmentation without fine-tuning.

We propose SegCLR as a CL solution to improve segmentation quality by leveraging both manual annotations and unlabeled data.
To that end we introduce a semi-supervised framework for joint training of CL together with segmentation labels.
Preliminary results of this work have been presented as a conference contribution~\cite{segclr} for slice-wise segmentation of two datasets acquired with different OCT scanners, where the domain shift originates from a change in acquisition device.
SegCLR without any target-domain labels is shown to achieve unsupervised domain adaptation results close to an upper bound that uses supervised learning on the target domain.
Our additional contributions in this work are:
(1)~In addition to acquisition device change, we present results with a third dataset with a new disease indication, which helps demonstrate the benefits of SegCLR also for domain shift in disease type.
(2)~An analysis of the amount of unlabeled target data is presented, indicating its limited effect on the resulting segmentation performance.     
(3)~Exploiting the finding above, we propose a domain generalization use-case of our SegCLR framework, for which an extensive analysis reveals substantial benefits showing that SegCLR trained models are implicitly better at generalizing to unseen domains. 
(4)~An analysis of the stability of the results upon distinct network initialisations reveals that most proposed SegCLR variants perform well and similarly, confirming the stability of the overall SegCLR framework to method setting choices such as pair generation and contrastive projection strategies. This further highlights the importance of analyzing many randomized result replicates in assessing deep learning models in general. 
(5)~In a multi-domain training scenario of SegCLR, we show that with labeled data available from multiple domains, SegCLR applied to the domains simultaneously yields results superior to conventional supervised training as well as SegCLR trained on individual domains, potentially better leveraging cross-domain information.

\section{Methods}
The considered learning frameworks are illustrated in Figure~\ref{fig:learning_frameworks}.
Let a \emph{source} domain ($D^s$) contain labeled data, with which \emph{Baseline} models can be trained following typical \emph{supervised} learning schemes.
Distinct domains for which only unlabeled images can be observed are considered \emph{target} domains ($D^t$). 
We first study an unsupervised domain adaptation setting as in~\cite{segclr}, where SegCLR is trained for a source domain $D^s$ and a target domain $D^t$ by using a combination of supervision and contrastive losses.
We then investigate its domain generalization adaptation application, where SegCLR is trained using only data from a labeled source domain $D^s$, while being applied on unseen target domains $D^t$, \ie on domains for which no data (labeled or unlabeled) is used in training. 

\begin{figure}
\centering
\includegraphics[width=\columnwidth]{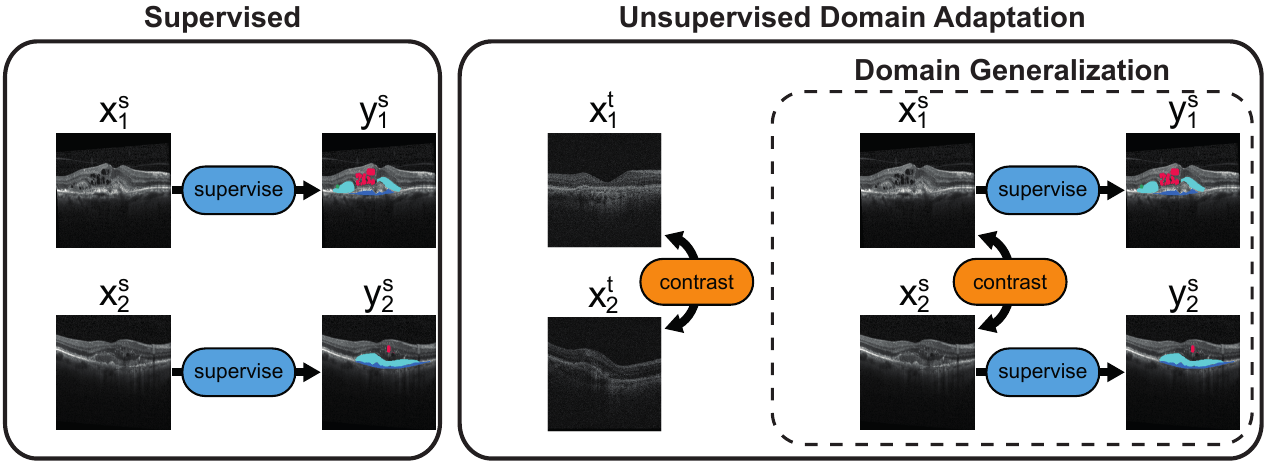}
\caption{Illustration of Unsupervised Domain Adaptation (UDA) and Domain Generalization  frameworks studied herein for SegCLR.
The colored ellipses indicate the losses to use in training, with the variable superscripts representing the domain being source (s) or target (t).}
\label{fig:learning_frameworks}
\end{figure}

\subsection{Supervised learning}
For the segmentation backbone, we adopted the proven UNet architecture~\cite{unet} (details in Figure~\ \ref{fig:network_scheme}), which can be modeled as $F(\cdot)$ processing an image $x$ to predict a segmentation map $p=F(x)$ to approximate an (expert-annotated) ground truth segmentation~$y$.
Each segmentation map contains $C$ classes, for which $y^c$ and $p^c$ represent, respectively, the ground truth and predicted probabilities of class $c$. 
The model $F$ learns to segment all classes present in the image by minimizing a supervised loss $\mathcal{L}_\mathrm{sup}$, which in our work is the logarithmic Dice loss of labeled data in a source domain $D^s$, \ie
\begin{equation}
    \mathcal{L}_\mathrm{sup} = -\frac{1}{C}\sum_{c=1}^{C}\sum_{(p_i,y_i) \in D^s}\log{\frac{2 \sum_{j \in \mathrm{pixels}} y_i^c p_i^c}{\epsilon + \sum_{j \in \mathrm{pixels}}  (y_i^c + p_i^c)}}
\end{equation}
for all training images $(x_i, y_i) \in D^s$, where $\epsilon$ is a small number ($10^{-12}$) to avoid division by zero.
Note that volumetric OCT data is processed slice-wise for 2D segmentation.

\begin{figure}[!t]
\centering
\includegraphics[width=\columnwidth]{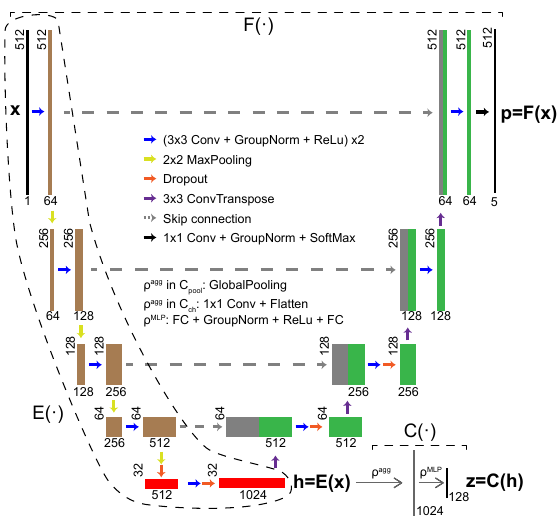}
\caption{
SegCLR architecture employed for joint supervised and contrastive learning.
Layers are represented as arrows and their outputs as rectangles.
The width and height of these outputs is annotated at the upper left of the rectangles, and the number of features at the bottom. 
$F(\cdot)$ is the segmentation backbone, $E(\cdot)$ the encoder, and $C(\cdot)$ the contrastive projection. 
The architectures of $\rho^\mathrm{agg}$ and $\rho^\mathrm{MLP}$ are described in Section~\ref{sec:contrastive_projection}.
}
\label{fig:network_scheme}
\end{figure}

\subsection{Self-supervised learning}
Self-supervised learning aims to optimize an encoder $E(\cdot)$ to achieve good representations $h=E(x)$ without the need of manually annotated labels~$y$. 
In the literature (for mostly classification tasks), the ResNet architecture is commonly used for $E(\cdot)$.
Instead, we use the existing UNet encoder in order to adapt the learned features $h$ for our intended segmentation task.
A subsequent contrastive projection head $C(\cdot)$ then maps the bottleneck-layer features to vector projections $z=C(h)$ on which the contrastive loss $\mathcal{L}_\mathrm{con}$ is applied. 
The architecture used for $E(\cdot)$ and $C(\cdot)$ is illustrated in Figure~\ref{fig:network_scheme}.
We herein employ two widely adopted self-supervised learning frameworks: SimCLR~\cite{simclr} and SimSiam~\cite{chen2021exploring}.

In SimCLR, $\mathcal{L}_\mathrm{con}$ aims to minimize the distance between \emph{positive} pairs of images and maximize the distance  between \emph{negative} pairs.
The positives $(x_i',x_i'')$ are created from each image $x_i$ by a defined pair generator $P(\cdot)$ described further in Section~\ref{sec:methods_pairing} below, i.e.\ $P(x_i) \rightarrow (x_i',x_i'')$. 
The negatives pairs $(x_i,x_k)$ can be formed using other images $x_k$ for $k \neq i$.
We employ a version of the normalized temperature-scaled cross entropy loss~\cite{oord2018representation} adapted to our problem setting as:
\begin{equation}
    L_\mathrm{con}^\mathrm{CLR}= \sum_{P(x_i),\ x_i \in D}\big(\,l(z_i', z_i'') + l(z_i'', z_i')\,\big)
\end{equation}
\begin{equation}
    l(z_i',z_i'') = -\log \frac{\exp{\big(d(z_i',z_i'')}/\tau\big)}{\sum_{x_k \in D} \mathbbm{1} _{[k \neq i]}\exp{\big(d(z_i',z_k)/\tau\big)}}
\end{equation}
where $d(u,v) = (u \cdot v) /  (||u||_2\, ||v||_2)$ and $\tau$ is the temperature scaling parameter. 

In SimSiam, a learnable predictor $Q(\cdot)$ is applied on the projection from one network path to predict that from the other:
\begin{equation}
     L_\mathrm{con}^\mathrm{Siam} = -\sum_{x_i \in D}\Big(d\big(Q(z_i'), z_i''\big) + d\big(Q(z_i''), z_i'\big)\Big)
\end{equation}
where the gradients from the second path are omitted for network weight updates (\emph{stopgrad}) in back-propagation.

\subsection{Joint semi-supervised and contrastive learning}\label{sec:segclr}
We adapt the USCL joint training strategy, which was proposed for US video classification, to our segmentation task by combining $\mathcal{L}_\mathrm{sup}$ and $\mathcal{L}_\mathrm{con}$ in a semi-supervised framework illustrated in Figure~\ref{fig:illustration_network}.
Considering a source domain $D^\mathrm{s}$ and a target domain $D^\mathrm{t}$, total loss $\mathcal{L}$ is calculated as follows:
\begin{equation}
    \mathcal{L} = \frac{1}{2} \left( \underset{x \in D^\mathrm{s}}{\mathcal{L}_\mathrm{con}} + \underset{x \in D^\mathrm{t}}{\mathcal{L}_\mathrm{con}} \right) + \lambda \underset{(x,y) \in D^\mathrm{s}}{\mathcal{L}_\mathrm{sup}}
\end{equation}
where $\lambda$ is a hyperparameter that controls relative contributions.

\begin{figure}[!t]
\centering
\includegraphics[width=\columnwidth]{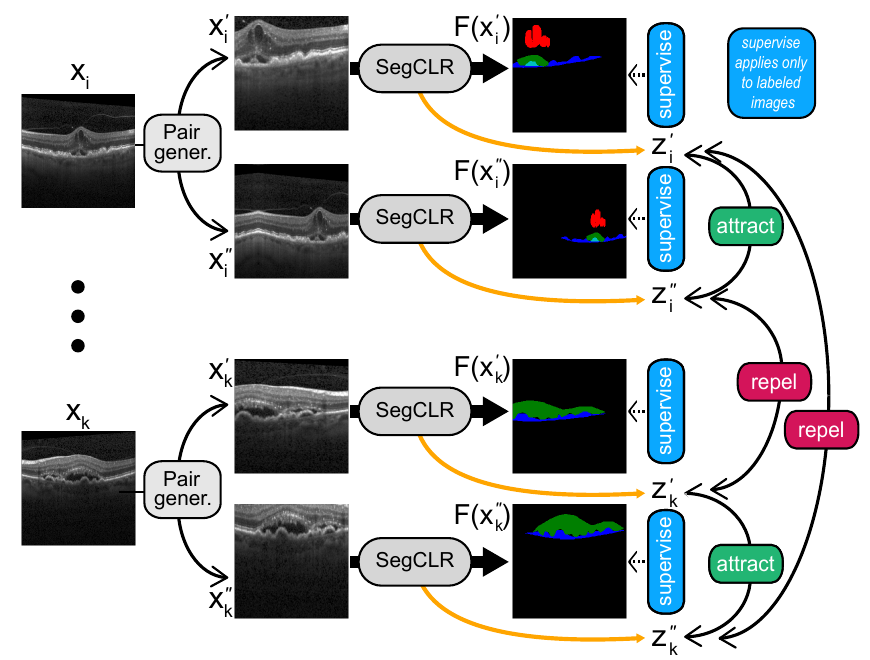}
\caption{
Illustration of semi-supervised contrastive learning framework for unsupervised domain adaptation. 
The SegCLR block corresponds to the architecture in Figure~\ref{fig:network_scheme}.
The \emph{repel} losses are not used by SimSiam.
Supervised losses are only used when labeled images exist. 
While this framework is flexible to accommodate any number of labeled images, at least one is required to drive the decoder arm of the underlying UNet. 
}
\label{fig:illustration_network}
\end{figure}

\subsection{Pair generation strategies}
\label{sec:methods_pairing}

The approach chosen for contrastive pair generation is key in self-supervised learning. 
We herein propose and compare different pair generation functions $P(\cdot)$ for volumetric OCT images, as illustrated in Figure~\ref{fig:pairing}.

First, the augmentation-based pair formation typically employed for natural images (\eg in SimCLR and SimSiam) is adapted for OCT slices, denoted as $P_\mathrm{a}$. 
To that end, labeled slices in $D^\mathrm{s}$ and random slices in $D^\mathrm{t}$ are augmented with horizontal flipping ($p=0.5$), horizontal and vertical translation (within 25\% of the image size), zoom in (up to 50\%), and color distortion (brightness up to 60\% and jittering up to 20\%). 
For color augmentation, images are first transformed to RGB and then back to grayscale.

Alternatively, we propose a slice-based pairing $P_\mathrm{s}$ that leverages the expected coherence of nearby slices in a 3D volume for CL.
For this, $x_i'=x_i$ is chosen with a slice index $b_i'$ in 3D. 
Then, $x_i''$ is a slice from the same volume with the (rounded) slice index $b_i''$ sampled from a Gaussian distribution $\phi$ centered on the index of the original image, \ie $b_i'' \sim \phi(b_i',\sigma)$, with standard deviation $\sigma$ as a hyperparameter. 
Combining the two pairing strategies yields $P_\mathrm{s+a}$ where $P_\mathrm{s}$ is used first and the augmentations in $P_\mathrm{a}$ are then applied on the selected slices.

\begin{figure}[!t]
\centering
\includegraphics[width=\columnwidth]{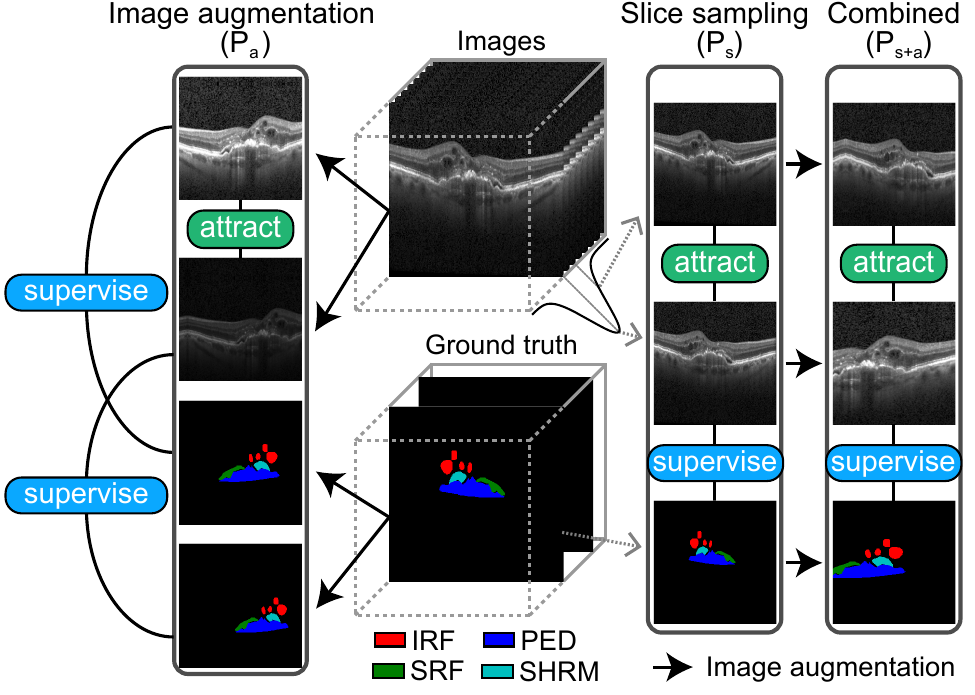}
\caption{Proposed pair generation approaches for contrastive learning with 3D images using SegCLR.}
\label{fig:pairing}
\end{figure}

\subsection{Aggregation of contrastive features}
\label{sec:contrastive_projection}
A projection head $C(\cdot)$ is formed by an aggregation function $\rho^\mathrm{agg}$ that aggregates features $h$ in a vector, which is then processed by a multilayer perceptron $\rho^\mathrm{MLP}$ to create projection~$z$. 
Typical contrastive learning frameworks such as SimCLR and SimSiam use a projection (denoted herein by $C_\mathrm{pool}$) where $\rho_\mathrm{pool}^\mathrm{agg}:  \mathbb{R}^{w \times h \times c} \rightarrow \mathbb{R}^{1 \times 1 \times c}$ is a global pooling operation on the width $w$, height $h$, and channels $c$ of the input features.
Such projection $C_\mathrm{pool}$ may be suboptimal for learning representations to effectively leverage segmentation information, as backpropagation from $\mathcal{L}_\mathrm{con}$ would lose the spatial context. 
Instead we propose $C_\mathrm{ch}$, for which $\rho_\mathrm{ch}^\mathrm{agg}: \mathbb{R}^{w \times h \times c} \rightarrow \mathbb{R}^{w \times h \times 1}$ is a $1$$\times$$1$$\times$$1$ convolutional layer that learns layer aggregation while preserving spatial context.

\subsection{Implementation}
Adam optimizer~\cite{kingma2014adam} was used in all models, with a learning rate of $10^{-3}$.
Dropout with \mbox{$p=0.5$} is applied on the layers illustrated with orange arrows in Figure~\ref{fig:network_scheme}. 
The projection head $\rho^\textrm{MLP}$ in $C(\cdot)$ is formed by two fully-connected layer with 128 units each, where the first one uses group normalization~\cite{wu2018group} (with a group size of 4) and ReLU activation.
We heuristically set $\lambda=20$ and the standard deviation for $P_\mathrm{s}$ as \mbox{$\sigma=250\,\mu m$}, which is the range for which we observe roughly similar features across slices.
Supervised models were trained for 200 epochs, and the model at the epoch with the highest average Dice coefficient across classes on the validation set was selected for evaluation on a holdout test set. 
Our implementation is in Tensorflow 2.7, ran on an NVIDIA V100 GPU. 

\subsection{Evaluation setup}\label{sec:methods_evaluation}
\label{sec:evaluation_strategies}
Individual slices of the 3D OCT volumes are segmented with UNet in 2D.
All reported experiments were replicated by training from scratch with 10 different initialization seeds, to factor out the sensitivity to randomness in network initialization.
%
Model performance was then also evaluated slice-wise and class-wise using slices with ground-truth annotations. The Dice coefficient serves as an overlap metric. In the situation that the ground truth and prediction maps are both empty, a perfect Dice score is given to that class on that slice. The Unnormalized Volume Dissimilarity (UVD) is also included as an error metric for the total segmentation error (FP+FN), \ie the sum of any (mis)segmented volume that is outside the ground truth and any ground truth volume that was not segmented by a method.
Compared to Dice, UVD is more robust to FP on  B-scans with small annotated regions for individual classes.
Dice is reported as \% (higher is better) and UVD as femtoliters (fL) (lower is better).   

Visualizing results by averaging metrics across classes with a large variation may lead to bias. Thus, for the metrics reported as boxplots, each per-slice metric $m^c_i$ for method $i$ and class $c$ is first normalized by its class Baseline $m^c_\textrm{bas}$, and then these are averaged over all $c$ and images on the test set.
The average of all Baseline replicates is used as $m^c_\textrm{bas}$ for calculating the \emph{relative metrics} of all other models.
Absolute metrics for each class are also reported in the tables as mean $\pm$ standard deviation.
To facilitate model comparison, the metrics included in most tables in the main text are averaged across classes.
To further allow for the interpretation and comparison of absolute segmentation performances, we included detailed tables in Supplementary Materials with metrics separately for individual labels.

Statistical significance is assessed using a paired t-test and  reported in the boxplots with **** for p-value$\leq$0.0001, *** for p-value$\leq$0.001, ** for p-value$\leq$0.01, * for p-value$\leq$0.05 (*), and `n.s.' for non-significant. When no bar joins two result boxes, the marker indicates significance for the test performed against the corresponding baseline. 

For comparing and ranking models, we adapted the strategy from~\cite{maier2017isles} to simultaneously consider both Dice and UVD as complementary metrics. 
Accordingly, for each individual volumetric eye image, the models are first ranked among themselves for each initialization seed, separately based on their Dice and UVD. 
These rankings are next averaged across the 10 seed replicates of all the volumetric images as well as across the two metrics, to obtain and report a final ranking of each model for a given test domain.

\section{Dataset}
\label{sec:dataset}
We employ three large OCT datasets from clinical trials.
As each dataset involve different eye diseases and/or acquisition devices, each then denotes a distinct domain $D_i, \, i \in \{1..3\}$, with details as follows:

\noindent$\bullet$ $D_1$: Images of neovascular age-related macular degeneration (nAMD) patients, acquired using a \emph{Spectralis} (Heidelberg Engineering) imaging device, yielding scans of $512 \times 496 \times 49$ or $768 \times 496 \times 19$ voxels, with a resolution of $10 \times 4 \times 111$ or $5 \times 4 \times 221$ $\mu$m/voxel, respectively.
    These were acquired as part of the phase-2 AVENUE trial (NCT02484690). 
    
\noindent$\bullet$ $D_2$: Images of nAMD patients, acquired using a \emph{Cirrus} HD-OCT III (Carl Zeiss Meditec) imaging device, yielding scans with $512 \times 1024 \times 128$ voxels and a resolution of $11.7 \times 2.0 \times 47.2$ $\mu$m/voxel.
    These were acquired as part of the phase-3 HARBOR trial (NCT00891735).

\noindent$\bullet$ $D_3$: Images of diabetic macular edema (DME) patients, acquired using a \emph{Spectralis} device with scan sizes and resolutions the same as for $D_1$.
    These were acquired as part of the phase-2 BOULEVARD trial (NCT02699450).

Selected B-scans from $D_1$ and $D_2$ were manually annotated for fluid regions of potential diagnostic value, \ie intraretinal fluid (IRF), subretinal fluid (SRF), pigment epithelial detachment (PED), and subretinal hyperreflective material (SHRM). 
B-scans from $D_3$ were annotated for IRF and SRF, but not PED nor SHRM, as these are not relevant for diabetic macular edema patients. 
More details on these datasets and the annotation protocol can be found in~\cite{maunz21}.

For computations in this work, all slices (B-scans) from the three datasets were resampled to $512 \times 512$ pixels using linear interpolation, yielding the approximate resolution of $10 \times 4$ $\mu$m/pixel. 
Corresponding ground-truth labels were also resampled accordingly using nearest neighbours. 

Labeled data exists for all 3 domains as detailed in \Cref{tab:dataset_details} together with the stratification details.
For each experimental setting as described in the corresponding following sections, different ablation combinations were performed by training models on labeled \emph{source} domain(s) $D^s$ and applying (testing) such trained models on unlabeled \emph{target} domain(s) $D^t$.
When a domain is considered $D^t$, we omit its labels in the training and use them only for ($i$)~evaluating the models on the test set of that domain, or ($ii$)~training of an \emph{UpperBound} model used as a reference.
The train/test configuration of each subsequent section is also summarized in \Cref{tab:experiment_details}.

\section{Results and Discussion}
In most experimental settings, we train a model $F$ concurrently on $(x,y) \in D^\mathrm{s}$ and $x \in D^\mathrm{t}$, and then apply this trained model on the target data for evaluating $F(x\,|\,x\!\in\!D^\mathrm{t})\rightarrow y \in D^\mathrm{t}$.
In the domain generalization setting, we train only on $(x,y) \in D^\mathrm{s}$, without any information from $D^\mathrm{t}$.
We evaluate the trained models also on their initial source domain $F(x\,|\,x\!\in\!D^\mathrm{s})\rightarrow y \in D^\mathrm{s}$ to assess the retention of source-domain segmentation capability.

\subsection{Joint learning for unsupervised domain adaptation across imaging devices }
\label{sec:avenue_harbor}
We first evaluate the proposed SegCLR framework in an unsupervised domain adaptation setting where 
the domain shift is caused by images being acquired with different imaging devices.
To address the DA between Spectralis and Cirrus devices, the source and target domains were chosen as $D^s=D_1$ and $D^t=D_2$, since the latter domain has many more unlabeled images to facilitate unsupervised DA.
This also helps us replicate the setting in~\cite{segclr} showing our preliminary results.

A supervised UNet model, \emph{Baseline}, is trained only on the source domain $D^\mathrm{s}$. 
An \emph{UpperBound} supervised model trained on labeled data from $D^t$ is included for comparison. 
This labeled data is used here only as a reference and is ablated for all other models. 
\Cref{tab:avenue_harbor} shows very poor results for Baseline on $D^t$, while results are much better for UpperBound on $D^t$ or Baseline on $D^s$, confirming that the two domains indeed differ from supervised learning perspective.
For easier comparison, \Cref{fig:metrics_avenue_harbor} shows the metrics relative to the average of Baseline models and averaged across labels, \ie so-called relative metrics.
The distribution across 10 training repetitions are shown to highlight the large inherent variation in these models, including Baseline.

For state-of-the-art comparison two additional approaches that does not utilize self-supervised learning are included.
CycleGAN~\cite{seebock2019using} was adapted to our UNet for OCT slices and it was trained to translate images of $D^t$ to the domain of $D^s$.
On such translated images, we then applied the Baseline UNet trained on $D^s$.
We qualitatively confirmed the convergence of training that yields meaningful translated images.
In \Cref{fig:metrics_avenue_harbor} the CycleGAN results for $D^t$ are seen to be slightly inferior to Baseline, contrary to the observation reported in~\cite{seebock2019using}. 
This is probably due to their baseline results having a near-zero Dice, while our Baseline being a relatively much superior contender (see \Cref{tab:avenue_harbor}). 
We also include singular value decomposition-based noise adaptation (SVDNA)~\cite{koch2022noise}, which was published concurrently with our preliminary results of SegCLR~\cite{segclr}. 
SVDNA is a style transfer technique that utilizes difference in noise structure between domains. 
We used its public-domain implementation~\cite{svdna_code}, by adapting its inherent parameter $k$, which denotes the threshold of singular values used, as $k \in [40,100]$ to accommodate for our different image size. 
The results for SVDNA in \Cref{fig:metrics_avenue_harbor} are superior to Baseline in the target domain.
Note that such methods for image/style transfer to a target domain are not intended for applying on the source domain, which is corroborated by the SVDNA results in \Cref{fig:metrics_avenue_harbor} having similar UVD and lower Dice compared to Baseline for $D^s$. 
The CycleGAN model is outright not applicable for the source domain.

Standard self-supervised learning methods learn the representation of $D^\mathrm{t}$.
Finetuning this subsequently on $D^\mathrm{s}$, SimCLR and SimSiam achieve a slight improvement over Baseline for $D^t$, as seen in \Cref{fig:metrics_avenue_harbor} by their ($\approx$2\%) higher Dice and lower UVD. 
This confirms that these pretraining strategies apply also on our OCT data.
The results for $D^s$ are comparable to Baseline, as the finetuning on this domain, \ie initializing the Baseline with the target-learned representations, does not seem to add more information than existing from the source supervision. 
Note that for SimCLR and SimSiam we use their standard implementations including their original sample pairing strategy (corresponding to $P_\mathrm{a}$ in our terminology) and contrastive-head aggregation approach (equivalent to $C_\mathrm{pool}$).

Our proposed \emph{SegCLR} framework is used here for joint training, which augments SimCLR with a supervised loss and for contrasting segmentations (Section~\ref{sec:segclr}).
SegCLR increases the number of network parameters only for training and merely by 6.85\% with respect to Baseline (UNet). 
The results in Figure~\ref{fig:metrics_avenue_harbor} show that most choices of the projection head $P$ and the contrastive head $C$ lead to much better results (higher Dice and lower UVD) for SegCLR, notably even surpassing the UpperBound in some cases.
%
The only configurations for which the results do not follow this positive trend involve the slice-based pairing strategy $P_s$ alone, indicating that merely contrasting nearby slices does not facilitate extracting features useful for segmentation. 
While $P_s$ has the potential to be a complementary augmentation to $P_a$, it does not seem to be useful for learning informative features by itself.
This corroborates other works on augmentations~\cite{caron2021emerging} reporting cropping to be a crucial augmentation, which in our work is excluded when $P_a$ is not applied.
Indeed, most SegCLR configurations yield satisfactory results, with only SegCLR($P_\mathrm{s}$,$C_\mathrm{pool}$) failing substantially compared to Baseline, both for $D^t$ and $D^s$. 
This is likely because the pooling operation removes the spatial context between nearby and hence relatively similar slices, hence preventing the learning of relevant features from spatial correspondence. 
In comparison, SegCLR($P_\mathrm{s}$,$C_\mathrm{ch}$) can easily capture pixel-wise differences between nearby slices, thanks to $C_\mathrm{ch}$ being spatially resolved, so that relevant features can be learned even with the $P_\mathrm{s}$ pairing. 

In~\cite{segclr} with a similar experimental setting we reported SegCLR($P_\mathrm{s+a}$,$C_\mathrm{ch}$) as the superior SegCLR configuration. 
Our strategy herein with repetition of experiments with different initializations (10 random seeds) reveals a large dependence on initializations, where sometimes different conclusions may be reached for specific seeds. 
While such repeated evaluation strategy is computationally expensive, we hope this observation to serve as an incentive to standardize evaluation processes for generalizable conclusions. 

Most configurations of SegCLR lead to results superior to Baseline.
In \Cref{tab:avenue_harbor}, SegCLR($P_\mathrm{a}$,$C_\mathrm{ch}$) is seen to produce the best or the second best segmentation results in most cases for both $D^s$ and $D^t$ for DA across imaging devices.
In particular, this configuration is seen to preserve the source domain information most successfully, with the best metric in $D^s$ --- notably even compared to the supervised Baseline.
This shows that supervised information from the labeled domain is not forgotten (\eg as a model capacity trade-off when learning from the large unlabeled domain) and is indeed even enhanced and enriched with the added contrastive strategy.
This is a feature that previous UDA methods such as CycleGAN and SVDNA do not have, as they aim at application on a specific target domain.
Hence, SegCLR can be used for segmentation of both $D^s$ and $D^t$, replacing the conventional Baseline approaches for the labeled domains, while approaching UpperBound such that it obviates labeling efforts for scenarios with no training data is or can be available in a target domain.

\begin{table}[b]
\centering
\caption{
Absolute segmentation metrics averaged across classes in the unsupervised domain adaptation setting using $D^s=D_1$ and $D^t=D_2$. Results with the best (underlined) and the second best performance are in bold (excluding UpperBound).
}
\resizebox{\linewidth}{!}{%
\begin{tabular}{|l|r@{}l:r@{}l|r@{}l:r@{}l|}
\hline
\multicolumn{1}{|c|}{\multirow{2}{*}{Model}} & \multicolumn{4}{c}{Dice [\%]} & \multicolumn{4}{|c|}{UVD [fL]} \\
\cdashline{2-9}
& \multicolumn{2}{c:}{$D^t$} & \multicolumn{2}{c|}{$D^s$} & \multicolumn{2}{c:}{$D^t$} & \multicolumn{2}{c|}{$D^s$} \\
\hline

UNet (UpperBound)                          &  61.30$\pm$&4.34 &            N/A& &    961$\pm$&142 &           N/A& \\
UNet (Baseline)                            &  39.23$\pm$&8.18 &  67.14$\pm$&8.24 &   1647$\pm$&215 &    600$\pm$&232 \\
\hdashline
CycleGAN                                   &  36.84$\pm$&7.42 &            N/A& &   1685$\pm$&203 &           N/A& \\
SVDNA                                      &  58.19$\pm$&5.53 &  65.86$\pm$&8.02 &  \B \underline{1013}$\pm$&143 &    610$\pm$&243 \\
\hdashline
SimCLR                                     &  41.01$\pm$&7.32 &  66.70$\pm$&8.09 &   1614$\pm$&189 &    608$\pm$&240 \\
SimSiam                                    &  41.14$\pm$&7.86 &  67.40$\pm$&8.63 &   1598$\pm$&189 &    603$\pm$&244 \\
\hdashline
SegCLR($P_\mathrm{a}$,$C_\mathrm{pool}$)   &  58.09$\pm$&5.85 &  67.24$\pm$&6.95 &   1177$\pm$&180 &    \B 598$\pm$&237 \\
SegCLR($P_\mathrm{s}$,$C_\mathrm{pool}$)   &  39.39$\pm$&8.38 &  52.24$\pm$&7.35 &   1725$\pm$&600 &    829$\pm$&327 \\
SegCLR($P_\mathrm{s+a}$,$C_\mathrm{pool}$) &  \B \underline{59.62}$\pm$&5.52 &  66.89$\pm$&7.32 &   1163$\pm$&216 &    597$\pm$&227 \\
SegCLR($P_\mathrm{a}$,$C_\mathrm{ch}$)     &  \B 58.32$\pm$&5.34 &  \B \underline{67.93}$\pm$&6.91 &   1162$\pm$&240 &    \B \underline{585}$\pm$&227 \\
SegCLR($P_\mathrm{s}$,$C_\mathrm{ch}$)     &  50.60$\pm$&8.98 &  66.38$\pm$&7.06 &   1325$\pm$&370 &    613$\pm$&241 \\
SegCLR($P_\mathrm{s+a}$,$C_\mathrm{ch}$)   &  58.09$\pm$&5.71 &  \B 67.29$\pm$&6.64 &   \B 1156$\pm$&203 &    607$\pm$&239 \\
\hline
\end{tabular}
}
\label{tab:avenue_harbor}
\end{table}

\begin{figure}[t]
\centering
\includegraphics[width=\columnwidth]{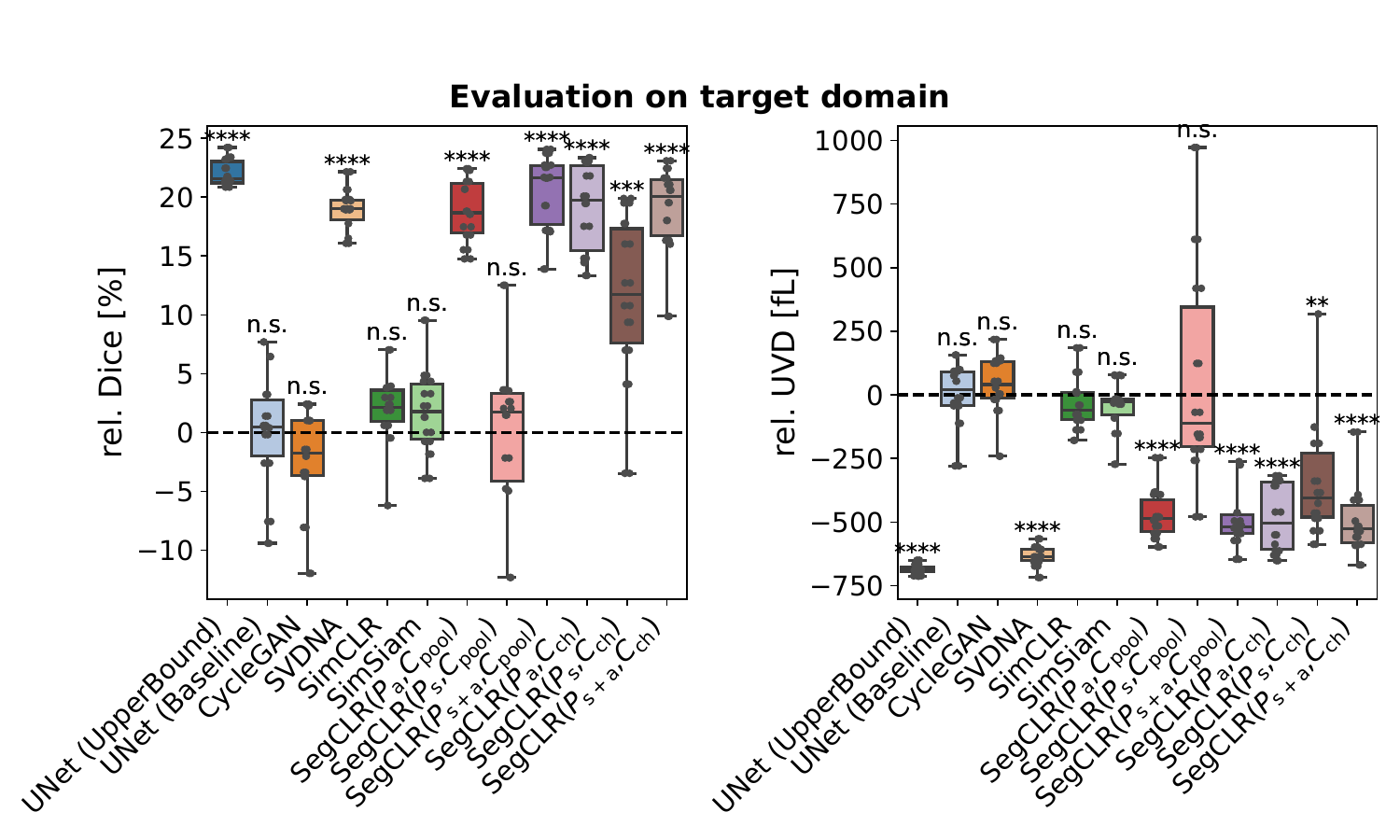}
\includegraphics[width=\columnwidth]{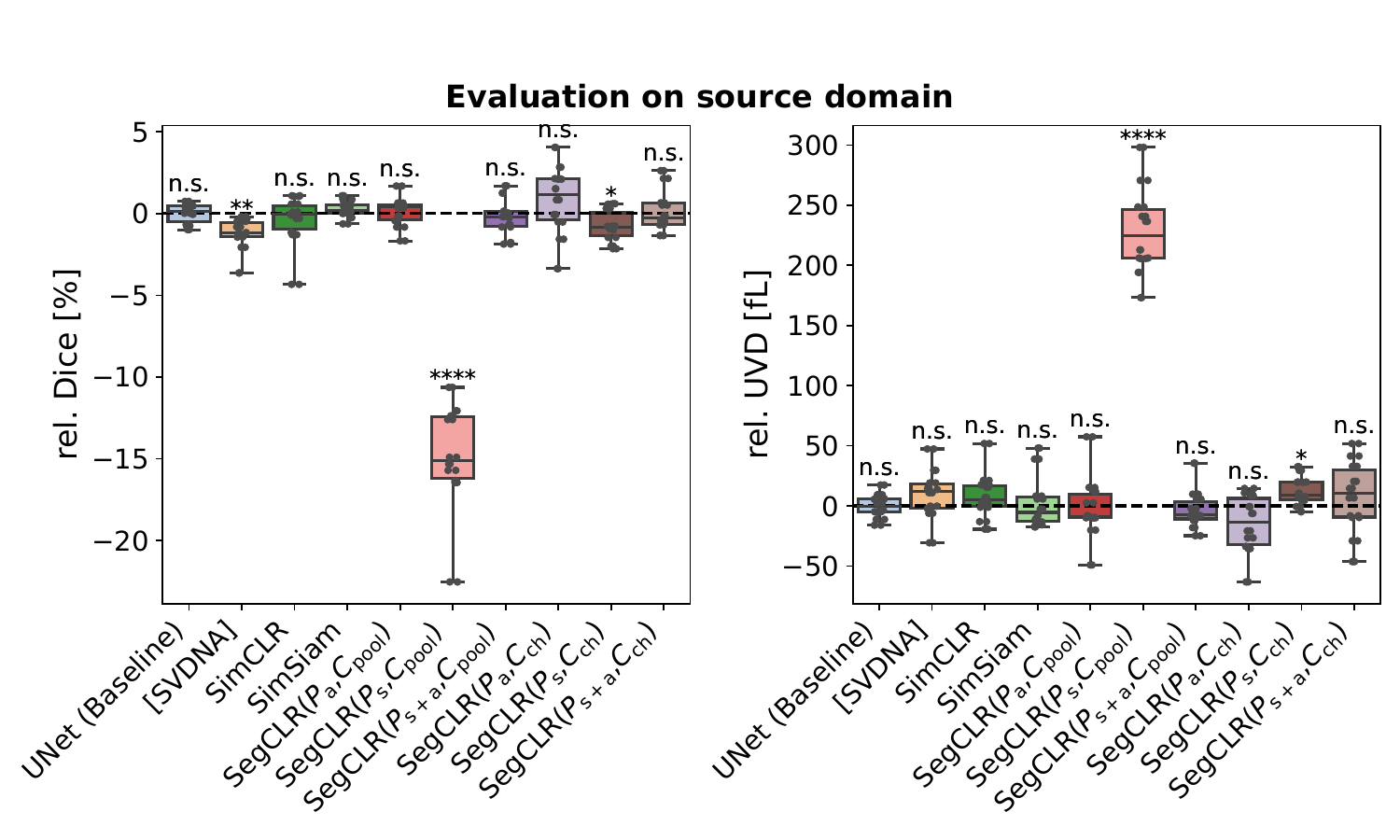}
\caption{
Relative segmentation metrics for cross-device domain adaptation, \ie using $D^s=D_1$ and ${D^t=D_2}$.
The dashed line depicts the average Baseline result used as reference for the relative metrics. 
SVDNA is included in brackets for $D^s$, as in practice Baseline model would be used instead.
}
\label{fig:metrics_avenue_harbor}
\end{figure}
 
\subsection{Unsupervised domain adaptation across eye diseases}
\label{sec:avenue_boulevard}
This section evaluates SegCLR for domain adaptation across eye diseases, \ie when image appearances are similar but the content differs.
To that end, an experimental setting with $D^s=D_1$ and $D^t=D_3$ is used.
Similarly to the previous section, \Cref{fig:metrics_avenue_boulevard} and \Cref{tab:avenue_boulevard} show that the Baseline model trained on $D^s$ performs substantially worse on $D^t$ compared to an UpperBound model trained on that target domain. 
As in the previous subsection, CycleGAN results are similar to Baseline. 
In contrast, SVDNA performs only slightly better than Baseline in the target domain, while it was clearly superior in the previous section. 
This is in line with the emphasis in~\cite{koch2022noise} that SVDNA as a noise adaptation method is successful mainly when the styles differ, \eg between different devices as in the previous subsection, whereas it seems not effective in scenarios with similar appearances but different content between domains as in this subsection.
Interestingly, in contrast to the previous section, SimCLR, \ie a model with contrastively pretrained features, is worse than Baseline for both $D^s$ and $D^t$ in this setting.
This can be due to instabilities in the contrastive loss, which do not affect SegCLR thanks to the proposed joint training strategy.

Similarly to the DA setting in the previous subsection, \Cref{fig:metrics_avenue_boulevard} confirms that SegCLR with most of its configurations substantially outperforms Baseline for $D^t$, while also yielding marginally better results for $D^s$. 
Only the configurations with $P_\mathrm{s}$ alone leads to results inferior to Baseline, with poorest results with the SegCLR($P_\mathrm{s}$,$C_\mathrm{pool}$) configuration.
In particular, \Cref{tab:avenue_boulevard} shows SegCLR($P_\mathrm{s+a}$,$C_\mathrm{ch}$) overall performs the best, with SegCLR($P_\mathrm{a}$,$C_\mathrm{ch}$) being a close follower.
%
The proposed method SegCLR being successful also in this different DA setting between eye diseases also involving different label sets reinforces the conclusions for its proposed use in diverse DA settings.
Indeed, its superior performance also in the source domain indicates its utility as a general learning framework for either labeled or unlabeled data from multiple domains is concerned.

\begin{figure}[h]
\centering
\includegraphics[width=\columnwidth]{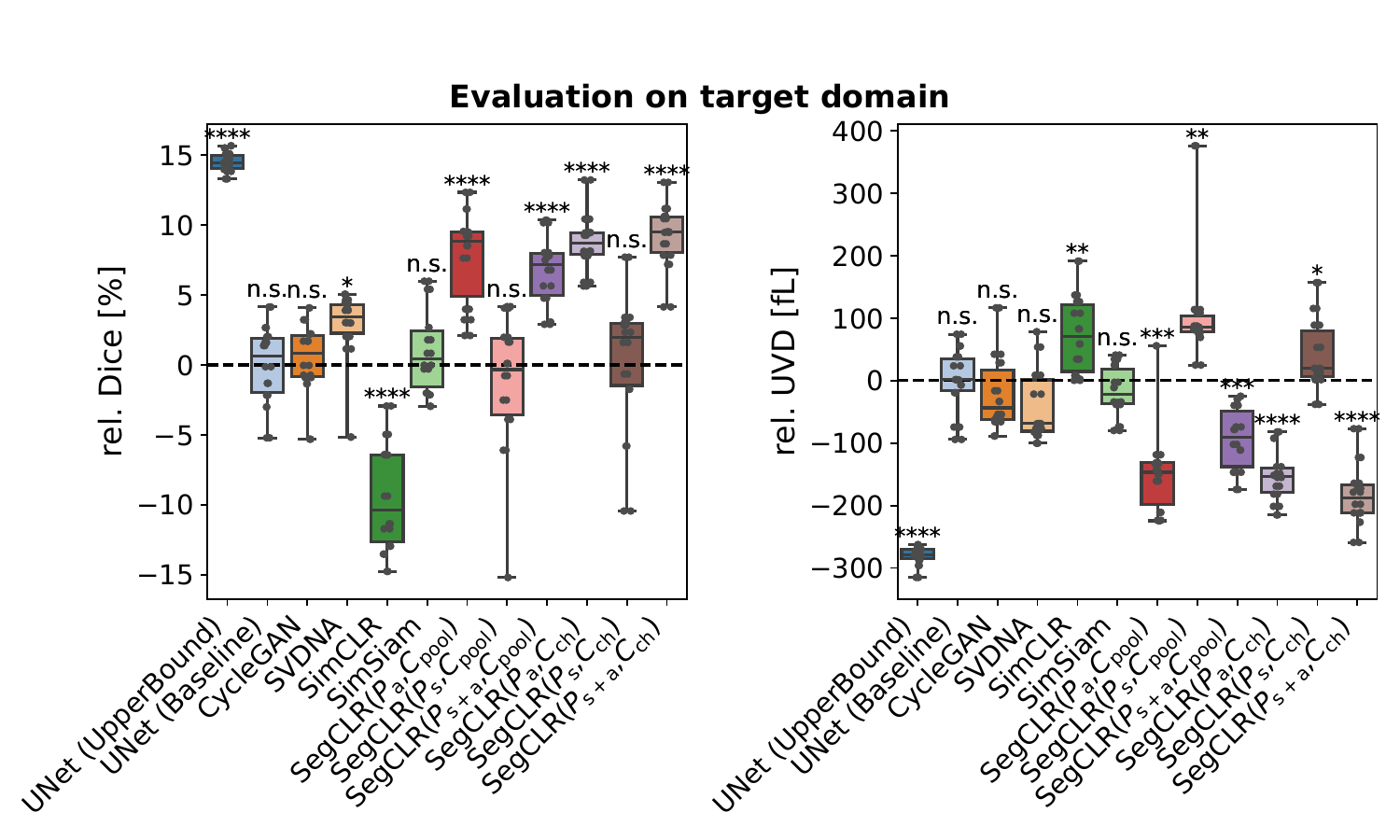}
\includegraphics[width=\columnwidth]{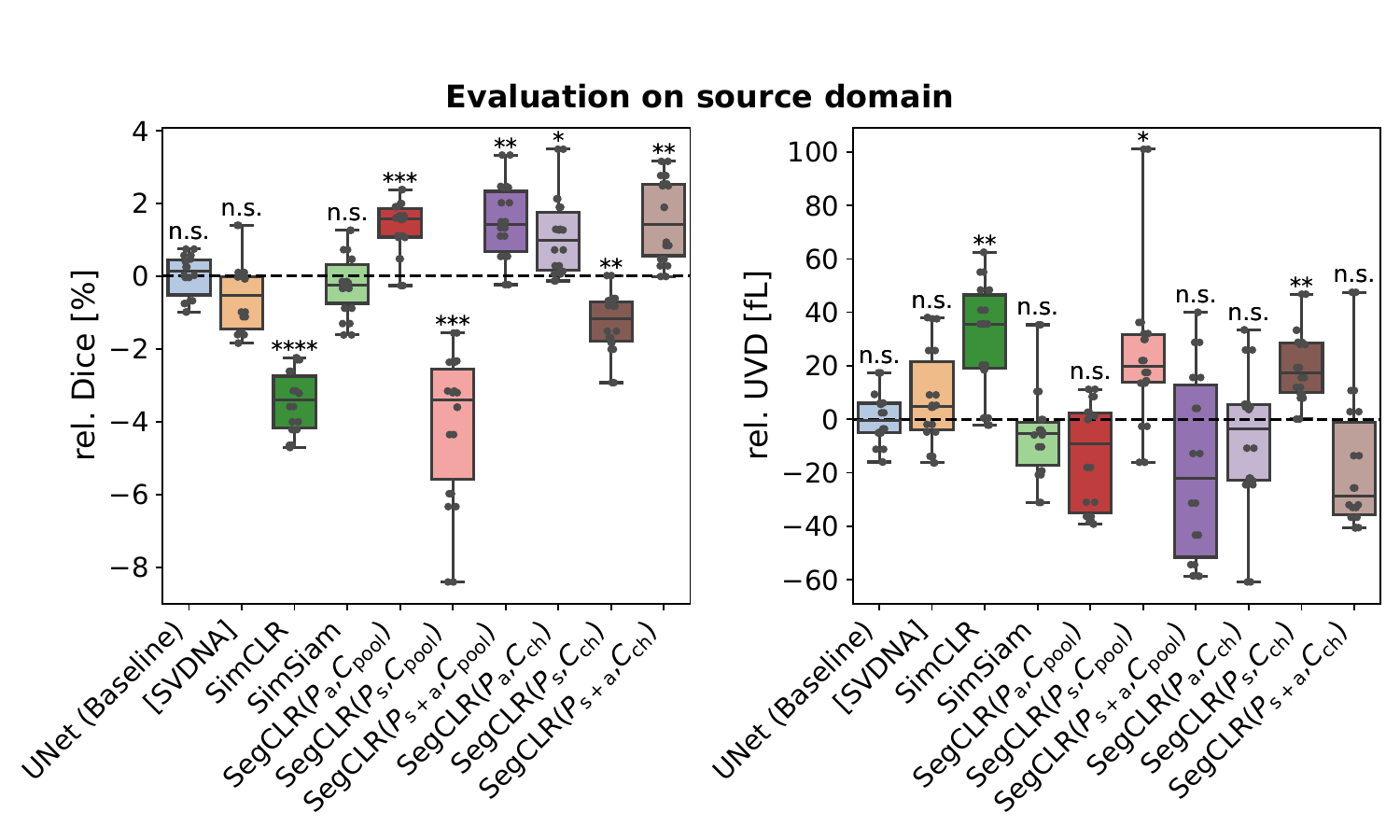}
\caption{
Relative segmentation metrics for cross-disease domain adaptation, \ie using $D^s=D_1$ and ${D^t=D_3}$.
The dashed line depicts the Baseline results used as reference for the relative metrics. 
SVDNA is included in brackets for $D^s$, as in practice Baseline model would be used instead.
}
\label{fig:metrics_avenue_boulevard}
\end{figure}

\begin{table}[b]
\centering
\caption{
Absolute segmentation metrics averaged across classes in the unsupervised domain adaptation setting using $D^s=D_1$ and $D^t=D_3$. Results with the best (underlined) and the second best performance are in bold (excluding UpperBound).
}
\resizebox{\linewidth}{!}{%
\begin{tabular}{|l|r@{}l:r@{}l|r@{}l:r@{}l|}
\hline
\multicolumn{1}{|c|}{\multirow{2}{*}{Model}} & \multicolumn{4}{c}{Dice [\%]} & \multicolumn{4}{|c|}{UVD [fL]} \\
\cdashline{2-9}
& \multicolumn{2}{c:}{$D^t$} & \multicolumn{2}{c|}{$D^s$} & \multicolumn{2}{c:}{$D^t$} & \multicolumn{2}{c|}{$D^s$} \\
\hline

UNet (UpperBound)                          &  79.32$\pm$&14.25 &            N/A& &    657$\pm$&491 &           N/A& \\
UNet (Baseline)                            &  64.77$\pm$&11.71 &  67.14$\pm$&8.24 &    938$\pm$&706 &    600$\pm$&232 \\
\hdashline
CycleGAN                                   &  65.24$\pm$&10.93 &            N/A& &    919$\pm$&695 &           N/A& \\
SVDNA                                      &  67.39$\pm$&12.14 &  66.56$\pm$&8.15 &    902$\pm$&698 &    608$\pm$&239 \\
\hdashline
SimCLR                                     &   55.33$\pm$&5.83 &  63.67$\pm$&6.53 &   1013$\pm$&670 &    632$\pm$&230 \\
SimSiam                                    &  65.71$\pm$&12.23 &  66.90$\pm$&7.81 &    921$\pm$&687 &    595$\pm$&233 \\
\hdashline
SegCLR($P_\mathrm{a}$,$C_\mathrm{pool}$)   &  72.49$\pm$&15.29 &  68.48$\pm$&7.36 &    794$\pm$&596 &    586$\pm$&231 \\
SegCLR($P_\mathrm{s}$,$C_\mathrm{pool}$)   &  63.13$\pm$&15.61 &  63.01$\pm$&7.41 &   1049$\pm$&708 &    625$\pm$&240 \\
SegCLR($P_\mathrm{s+a}$,$C_\mathrm{pool}$) &  71.48$\pm$&15.94 &  \B 68.64$\pm$&6.94 &    846$\pm$&587 &    \B \underline{583}$\pm$&233 \\
SegCLR($P_\mathrm{a}$,$C_\mathrm{ch}$)     &  \B 73.52$\pm$&16.07 &  68.25$\pm$&7.01 &    \B 785$\pm$&593 &    593$\pm$&241 \\
SegCLR($P_\mathrm{s}$,$C_\mathrm{ch}$)     &  65.00$\pm$&11.09 &  65.85$\pm$&7.10 &    982$\pm$&668 &    620$\pm$&250 \\
SegCLR($P_\mathrm{s+a}$,$C_\mathrm{ch}$)   &  \B \underline{73.99}$\pm$&16.40 &  \B \underline{68.67}$\pm$&7.45 &    \B \underline{756}$\pm$&575 &    \B 584$\pm$&225 \\
\hline
\end{tabular}
}
\label{tab:avenue_boulevard}
\end{table}

\subsection{Selecting the reference SegCLR implementation}
To select a specific SegCLR configuration for subsequent experiments, we tabulate average rankings of the models in \Cref{tab:ranking_uda} that summarize the results of both studied DA settings from the previous two subsections. 
\begin{table}[b]
\centering
\caption{
Ranking (lower is better) of method variants for the experiments of domain adaptation across devices in \Cref{sec:avenue_harbor} ($D^t=D_2$) and across diseases in \Cref{sec:avenue_boulevard} ($D^t=D_3$) following the reporting scheme described in \cref{sec:evaluation_strategies}. 
The results are sorted by Overall ranking, which is the average ranking from both experiments across $D^s$ and $D^t$. 
}
\resizebox{\linewidth}{!}{%
\begin{tabular}{|l|cc|cc|c|}
\hline
\multirow{2}{*}{Model} & \multicolumn{2}{c|}{$D^t=D_2$} &  \multicolumn{2}{c|}{$D^t=D_3$} & \multirow{2}{*}{Overall}  \\
\cdashline{2-5}
 & $D^t$ &  $D^s$ & $D^t$ &  $D^s$ &  \\
\hline
SegCLR($P_\mathrm{a}$,$C_\mathrm{ch}$) & 4.08 & 5.56 & 4.57 & 5.60 & 5.01 \\
SegCLR($P_\mathrm{s+a}$,$C_\mathrm{ch}$) & 4.11 & 5.96 & 4.39 & 5.45 & 5.03 \\
SegCLR($P_\mathrm{a}$,$C_\mathrm{pool}$) & 4.28 & 5.65 & 4.98 & 5.54 & 5.19 \\
SegCLR($P_\mathrm{s+a}$,$C_\mathrm{pool}$) & 4.00 & 5.97 & 5.29 & 5.50 & 5.32 \\
SVDNA & 4.59 & 5.41 & 5.89 & 5.83 & 5.55 \\
Baseline & 7.90 & 5.41 & 5.68 & 5.83 & 6.00 \\
SimSiam & 7.34 & 5.17 & 6.55 & 5.67 & 6.09 \\
SegCLR($P_\mathrm{s}$,$C_\mathrm{ch}$) & 5.66 & 5.91 & 6.52 & 6.42 & 6.20 \\
CycleGAN & 8.66 & 5.41 & 6.46 & 5.83 & 6.37 \\
SimCLR & 7.44 & 5.69 & 8.26 & 7.26 & 7.21 \\
SegCLR($P_\mathrm{s}$,$C_\mathrm{pool}$) & 7.93 & 9.86 & 7.40 & 7.08 & 8.04 \\
\hline
\end{tabular}
}
\label{tab:ranking_uda}
\end{table}

The rankings confirm the clear advantage of most SegCLR models relative to Baseline, SimCLR, and SimSiam. 
They also show that pretrained models do not help in such unsupervised domain adaptation tasks, for which joint training with SegCLR appears to be the key (with the only exception for the use of the pairing strategy $P_\mathrm{s}$, as discussed above).
For ranking of CycleGAN and SVDNA on $D^s$, the Baseline results are used in their stead, not to disadvantage these methods as this would also be the practical deployment scenario. 
As mentioned above, CycleGAN is not applicable to $D^s$. While SVDNA is applicable, it was designed for $D^t$ only, as confirmed by the results above in \Cref{fig:metrics_avenue_harbor} and \Cref{fig:metrics_avenue_boulevard}, where its performance is comparable to or worse than Baseline. 
While CycleGAN ranks (6.37) slightly worse than Baseline (6.00), the improvement of SVDNA on $D^t$ makes it rank slightly higher overall (5.55). 
Despite being a close contender, SVDNA ranks behind most SegCLR configurations for UDA, \ie in $D^t$ rankings including cross-device ($D^t=D_2$) setting, for which SVDNA was originally designed for.
While SVDNA can merely mirror Baseline results for $D^s$ evaluations, SegCLR can leverage target domain info back for the source domain, facilitating its success for $D^s$ in the cross-disease ($D^t=D_3$) setting.
Overall, most SegCLR configurations perform superior to SimSiam, SimCLR, CycleGAN, and SVDNA. 

Although the rankings differ slightly depending on whether the source or the target domain performance is prioritized, overall SegCLR($P_\mathrm{a}$,$C_\mathrm{ch}$) is seen to be the best performer. 
By not requiring slice sampling, this model also has a simpler implementation in practice, \eg compared to SegCLR($P_\mathrm{s+a}$,$C_\mathrm{ch}$), which is a top contender.
In addition, the proposed $C_\mathrm{ch}$ adds a mere 0.03\% more parameters to the baseline model.
Accordingly, we chose SegCLR($P_\mathrm{a}$,$C_\mathrm{ch}$) as the reference configuration for our proposed method, calling it simply SegCLR henceforth, to use in the subsequent experiments and ablations. 

\subsection{Effect of amount of target data on SegCLR}
\label{sec:unlabeled_ablation}
The success of modern self-supervised learning frameworks has been largely attributed to the availability and access to large unlabeled datasets~\cite{simclr,grill2020bootstrap,chen2021exploring,chen2020big}. 
Previous sections demonstrate the efficacy of SegCLR given such large datasets, from extensive clinical trials which are rarely available in practice.
An essential question thus remains as to how such framework performs in scenarios with smaller sets of unlabeled data.
This question is studied here as an ablation study by reducing the amount of unlabeled data on the target domain $D^t$.
We evaluate this with the experimental setting  for DA across imaging devices (\Cref{sec:avenue_harbor}), \ie $D^s=D_1$ and $D^t=D_2$.

\Cref{fig:unlabeled_ablation} presents the results of SegCLR when trained with logarithmically reducing amount of unlabeled data compared to the earlier experiments. 
The experiments are taken down to the point of including no unlabeled data, \ie the contrastive loss being computed only using the source domain data (jointly with its supervised loss).
Results show that the source domain performance is not affected much by the amount of unlabeled data from a different domain, which is expected since the supervised source domain is the major source of information and the main driver for this task.
Interestingly, however, regardless of the amount of unlabeled data, SegCLR performs roughly 1\% in Dice on average compared to Baseline.
This suggests that the proposed framework and contrastive strategy implicitly advances and improves standard UNet training and its generalization to test data from the same source domain, regardless of data existence from yet another domain.

Results in \Cref{fig:unlabeled_ablation} for the target domain of interest shows an expected trend of deterioration with diminishing amount of target data.
Nevertheless, surprisingly the reduction in performance is quite minor, and a reduction of only 5.25$\pm$7.33\% Dice and -0.99$\pm$2.13\% UVD is observed between no unlabeled target data and 100\%  unlabeled target data (of nearly 7 million B-scans).
Strikingly, SegCLR without any unlabeled data (\ie SegCLR for domain generalization) performs still significantly superior to Baseline, by 13.84$\pm$6.22\% Dice and -3.85$\pm$1.64\% UVD.
This again suggests that the introduced framework with the joint learning strategy inherently enables better generalization, even in supervised settings without requiring additional data and without adding much to the network complexity.
This is indeed a remarkable feature of our proposed method in contrast to previous UDA methods (\eg CycleGAN and SVDNA), which by design require unlabeled data from the target domain.
We further study this surprising observation in the next section.

\begin{figure}[!t]
\centering
\includegraphics[width=\columnwidth]{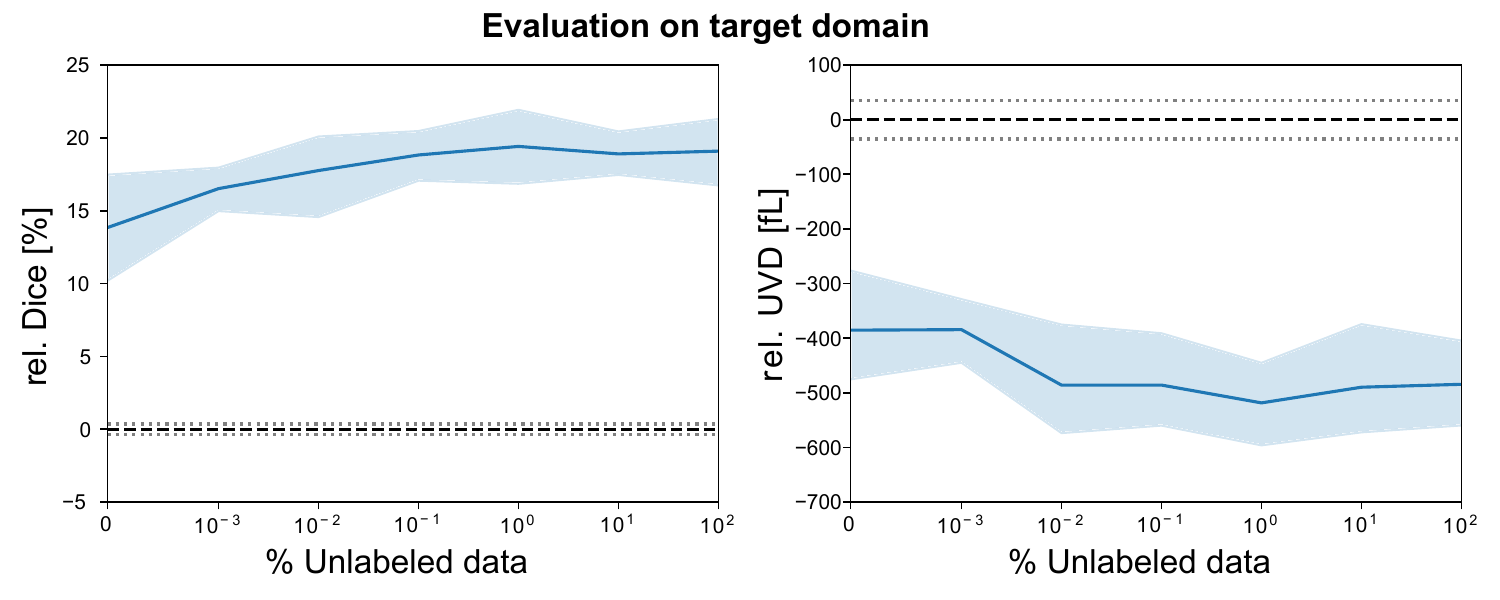}
\includegraphics[width=\columnwidth]{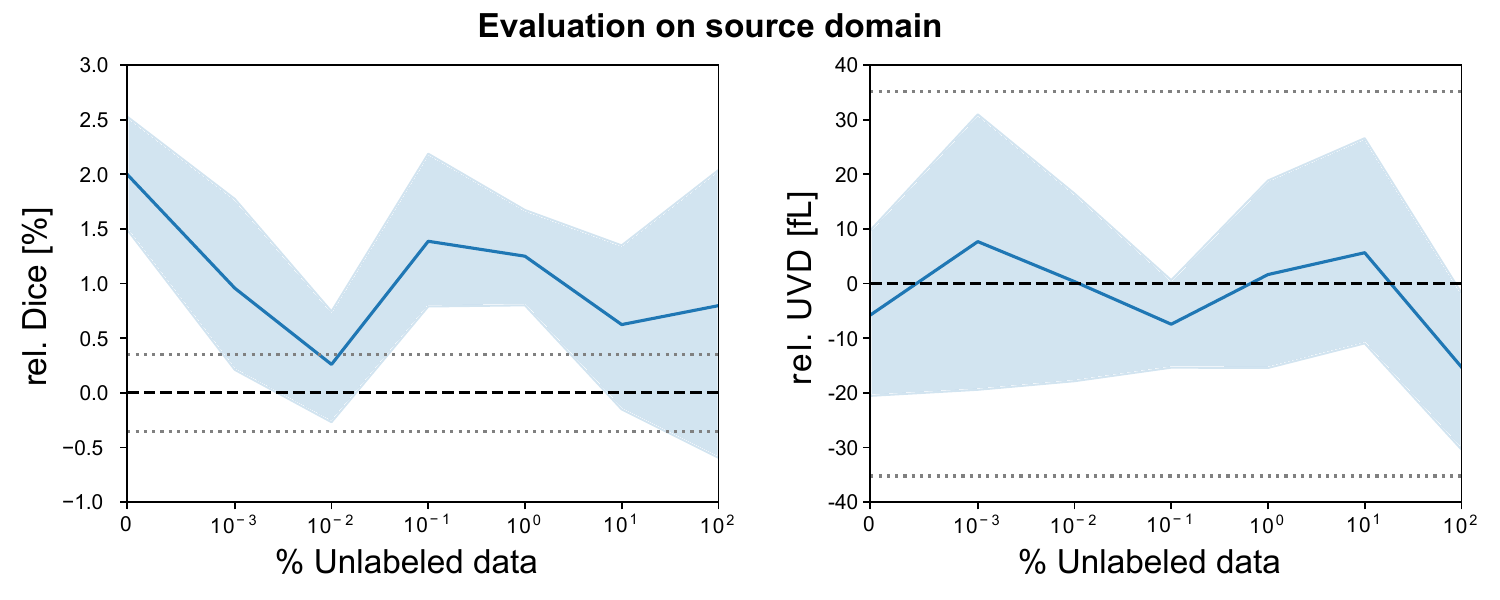}
\caption{
Segmentation results of SegCLR for ablation experiment, where unlabeled data on the target domain $D^t$ is systematically reduced. 
The 95\% confidence interval is shown as blue bands for SegCLR and as gray dotted lines for Baseline replicates.
The average of Baseline replicates (black dashed line at 0) is the basis for relative comparisons, \ie a rel.\,Dice above 0 or a rel.\,UVD below 0 indicates a superior performance by SegCLR.
Note the different y-axis scales of the source and target domain plots.
}
\label{fig:unlabeled_ablation}
\end{figure}

\subsection{Domain generalization on different domain combinations}
\label{sec:zero_shot}
In this section, we assess SegCLR domain generalization capability across different experimental settings.
To that end, we consider all possible combinations of the datasets $D_1$, $D_2$, and $D_3$ as $D^s$ and $D^t$.
Additionally, we perform evaluations on a combined dataset $D_\mathrm{All}=D_1 \cup D_2 \cup D_3$ as $D^s$, which is motivated and detailed further in the next section.
Note that in this domain generalization setting, $D^t$ is exclusively reserved for evaluation, with its data (including unlabeled images) not been seen during the training phase.

\Cref{fig:results_zeroshot} and \Cref{tab:results_zeroshot_all} demonstrate the superiority of SegCLR over UNet consistently in nearly all experimental dataset combinations.
The only setting SegCLR is clearly inferior (for both Dice and UVD) is for training on $D^s=D_3$ and evaluating on $D^t=D_1$, for which the imaging device stays the same but the disease type (labeling info and label sets) mainly differ.
Note that with the imaging device DA from $D^s\in\{D_1,D_2\}$ to $D^t=D_2$ leads a relatively larger reduction in performance for UNet, which cannot generalize across such large domain gap --- a setting in which SegCLR appears to be relatively successful.
Visual examples from different experimental combinations are shown in \Cref{fig:visuals_zs}.

\begin{figure}[t]
\centering
\includegraphics[width=\columnwidth]{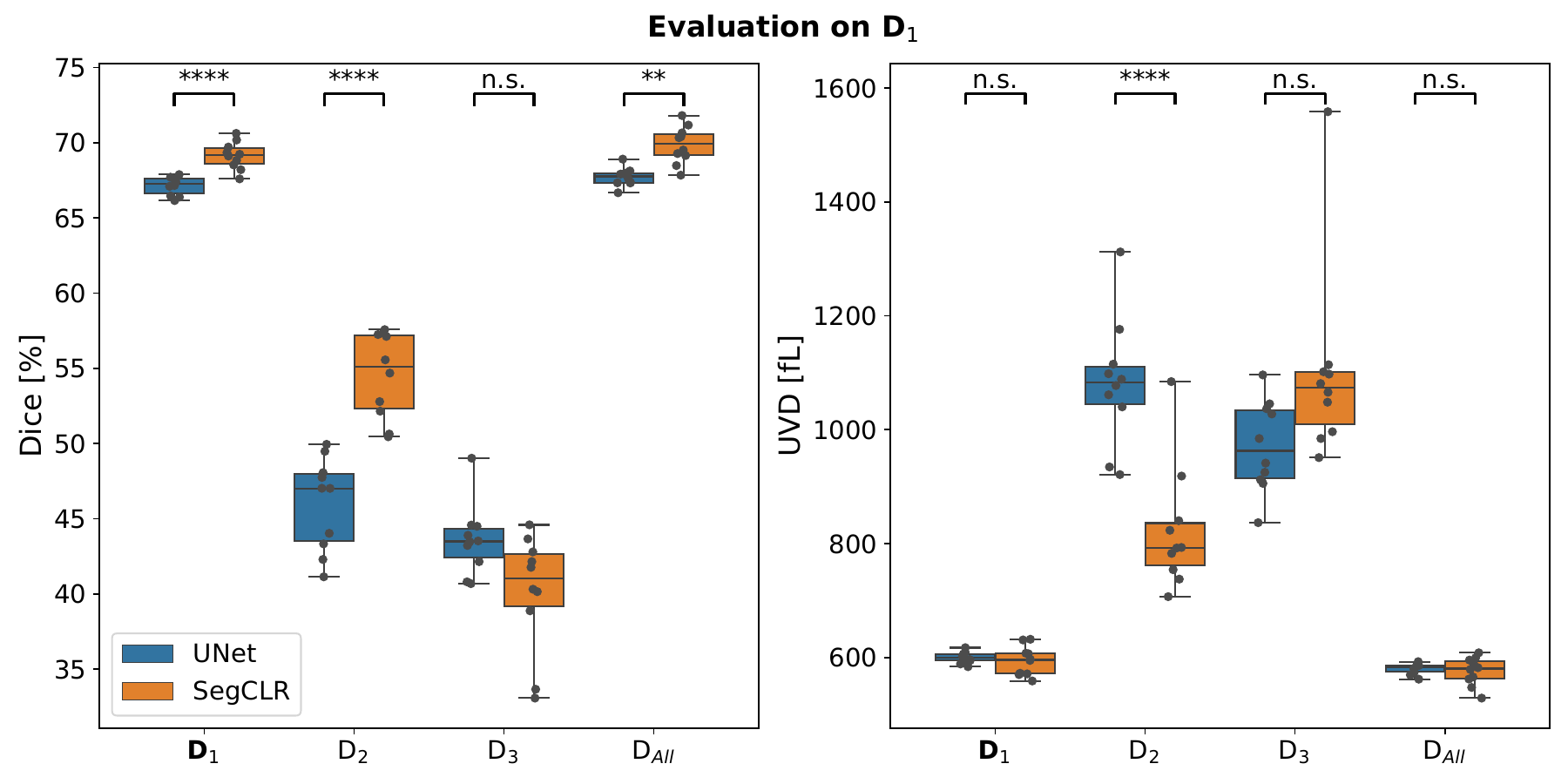}
\includegraphics[width=\columnwidth]{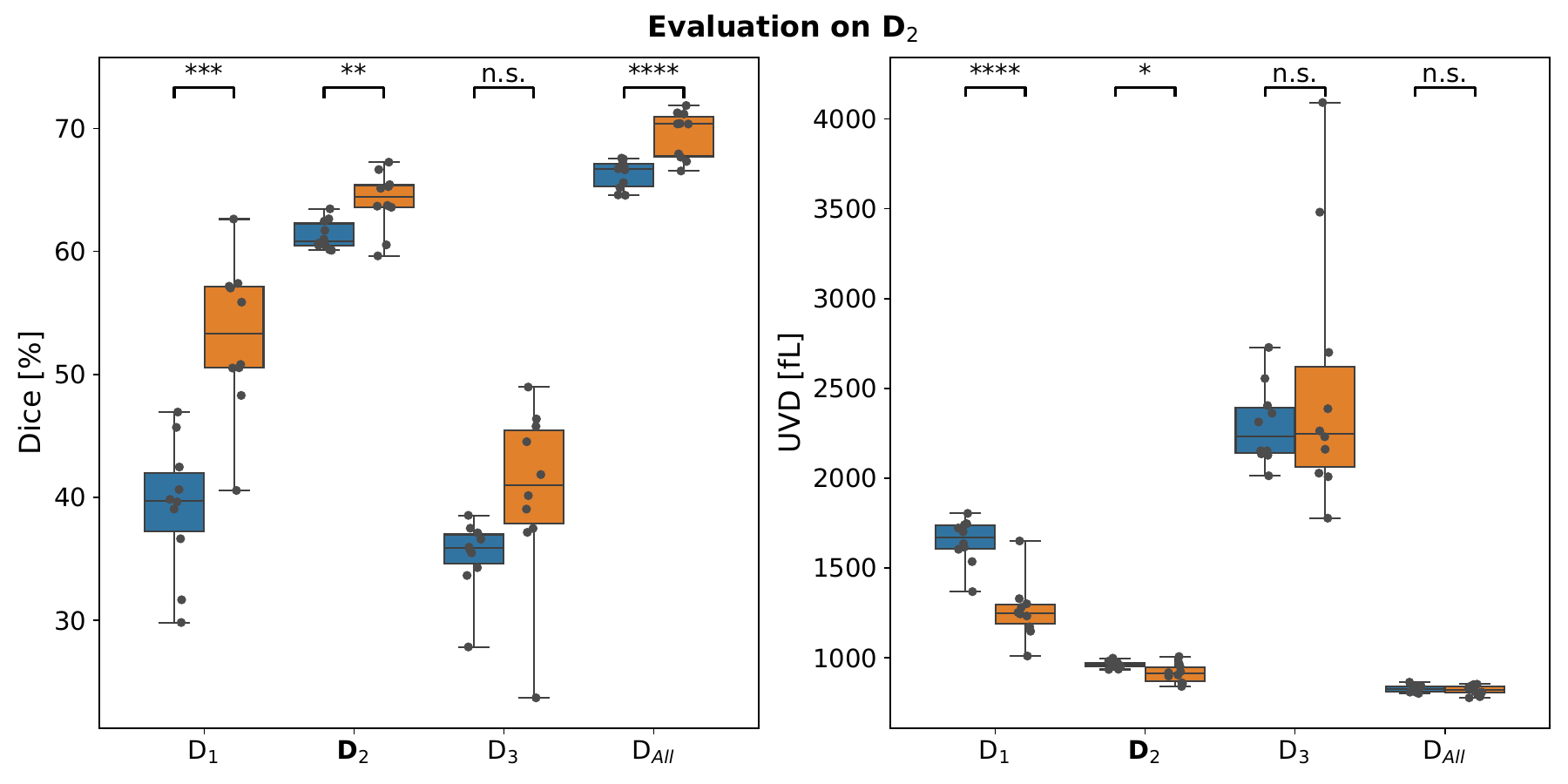}
\includegraphics[width=\columnwidth]{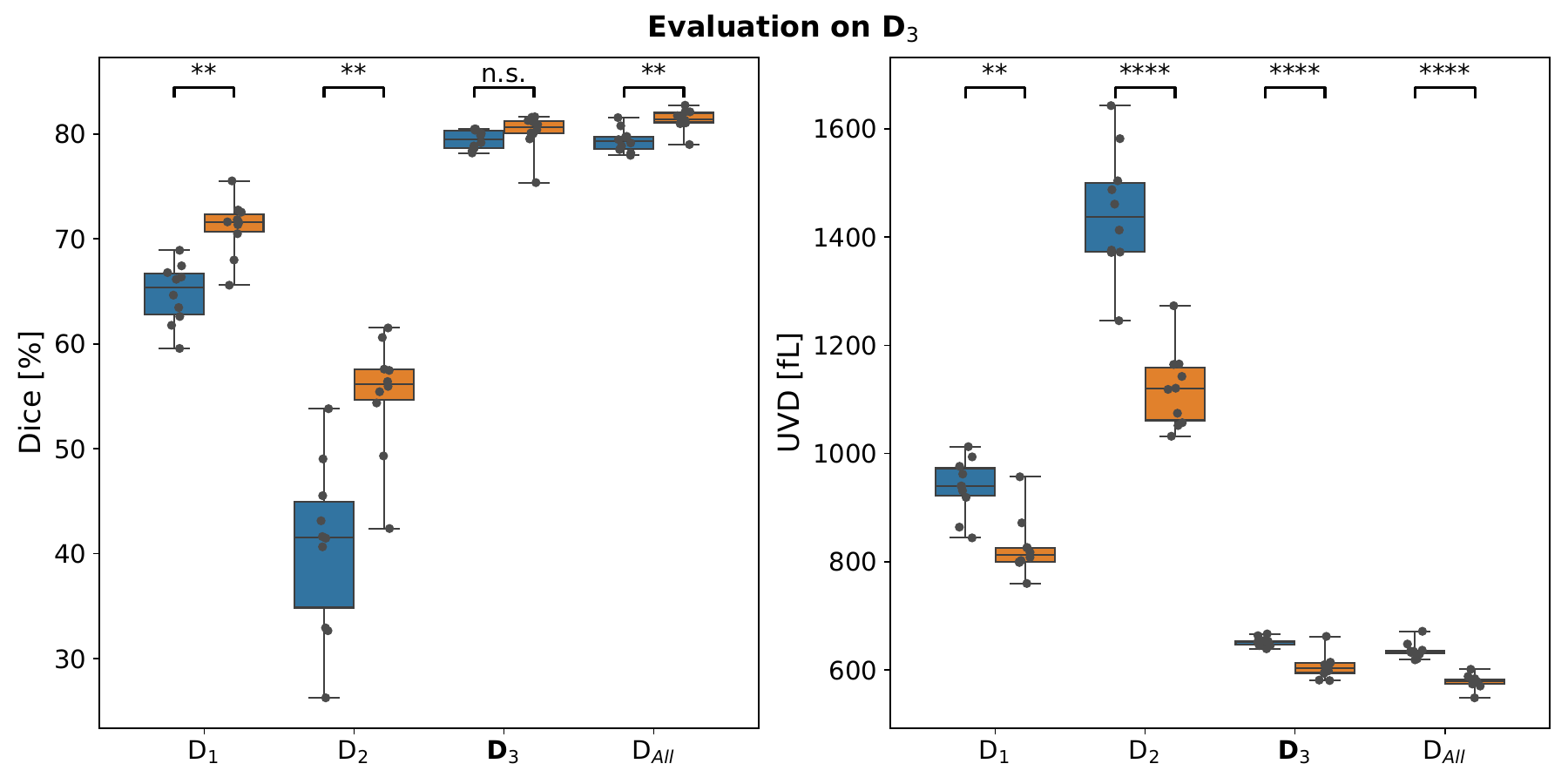}
\caption{
Results for domain generalization models trained on different source domains $D^s$ shown in the x axes, with $D_\mathrm{All}=D_1 \cup D_2 \cup D_3$.
Each row of plots shows the evaluation results on a target domain $D^t$. 
In the supervised learning setting when the evaluation domain matches the training domain (indicated by bold x-labels), UNet results correspond to UpperBound per our previous naming.
In other settings, these UNet results are the Baseline, \ie application without any specific domain adaptation effort. Although the margin for improvement becomes small with both models performing well in this supervised setting, SegCLR can surprisingly still outperform UNet in both Baseline and UpperBound settings. 
SegCLR can thus augment and replace conventional UNet for better generalizability.
}
\label{fig:results_zeroshot}
\end{figure}

Overall, SegCLR is seen to outperform the UNet baseline across different combinations of source and target datasets.
Additionally, SegCLR surpasses UNet even when evaluated within the same source domain, the scenario denoted as UNet (UpperBound) in the previous sections. 
This suggests that the benefits of SegCLR extend beyond DA alone, proving valuable even for conventional supervised settings with training and deployment within the same domain, \ie as a plug-and-play replacement for UNet.

Collectively, these results affirm the prior hypothesis that SegCLR exhibits robust performance in realistic scenarios where deployment on entirely unseen domains is anticipated, particularly when the domain gap arises from changes image appearance, \eg across different imaging devices.

\begin{figure}
\centering
\includegraphics[width=0.85\columnwidth]{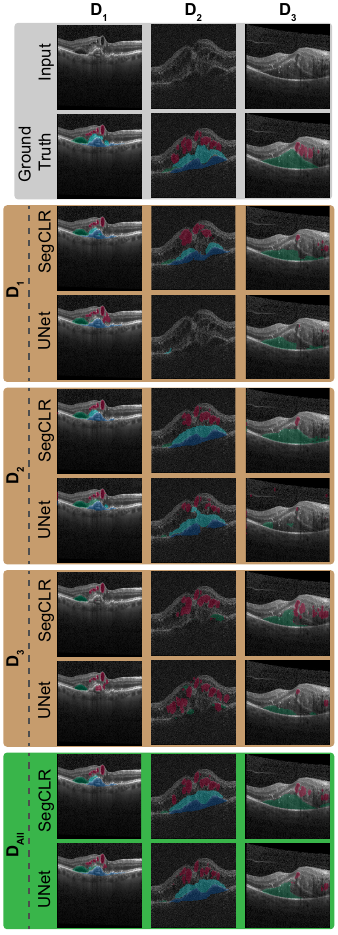}
\caption{Qualitative assessment of SegCLR and UNet baseline, trained on different domains (rows) for example images from each domain (columns). Note that $D_3$ only has two of the four labels. 
These visually exemplify that ($i$)~training and evaluating on the same domain works well with both UNet and SegCLR, ($ii$)~SegCLR outperforms UNet when training and evaluation domains differ, \ie in UDA settings, ($iii$)~multi-domain training $D_\mathrm{All}$ performs better than single-domain training, for both UNet and SegCLR, with minor superiority of SegCLR (see bottom-right for $D_3$).}
\label{fig:visuals_zs}
\end{figure}

\subsection{Multi-domain supervised contrastive learning}
\label{sec:multi_domain}
Although the evaluations so far above considered only single source domains, it is common in real-world contexts to have data available from multiple domains. 
For conventional supervised training, preliminary findings, \eg\cite{gomariz2022unified}, suggest that training with multi-domains simultaneously may yield superior segmentation outcomes on each domain separately, compared to models trained individually on each domain. 
This indicates that different domains may bring in complementary and supporting information in a multi-domain training setting.
We test here SegCLR in such practical setting of multi-domain training.

We train both UNet and SegCLR models on $D^s=D_\mathrm{All}\equiv D_1 \cup D_2 \cup D_3$, evaluating the two trained models on each domain separately.
Note that no unlabeled data is used in this setting, where only the images from the annotated training data were used for contrastive learning.
Quantitative results are shown in \Cref{fig:results_zeroshot} and \Cref{tab:results_zeroshot_all}, with examples depicted in \Cref{fig:visuals_zs}. 
The results corroborate the findings in~\cite{gomariz2022unified} in that UNet trained on $D_\mathrm{All}$ outperforms all other individual UNet models for the corresponding target domains (\ie comparing the rows of UNet across each column), except for a minuscule margin in Dice for $D_3$.
The observation further extends to SegCLR, \ie when trained on $D_\mathrm{All}$, SegCLR consistently outperforms UNet as well as all other individual-domain-trained SegCLR results, for all examined domains and both metrics. 
Evidently, SegCLR framework can effectively leverage contrastive information even from labeled data, thereby enhancing model generalizability across datasets as well as in-domain image variations. 
Consequently, SegCLR emerges as an ideal framework also for supervised learning only with labeled data, \ie when there is no domain shift.
This further includes settings involving multiple domains for training and/or inference.

\begin{table}
\centering
\caption{
Domain generalization results for models trained on different $D^s$ (rows) and evaluated on different $D^t$ (columns).
Absolute metrics are shown averaged across all classes, in bold for the best performance.
}
\resizebox{\columnwidth}{!}{%
\begin{tabular}{|c|c|r@{}lr@{}lr@{}l|r@{}lr@{}lr@{}l|}
\hline
\multicolumn{1}{|c|}{\multirow{2}{*}{$D^s$}} & \multicolumn{1}{c|}{\multirow{2}{*}{Model}} & \multicolumn{6}{c|}{Dice [\%]} & \multicolumn{6}{c|}{UVD [fL]} \\
\cdashline{3-14}
& & \multicolumn{2}{c}{$D_\textrm{1}$} & \multicolumn{2}{c}{$D_\textrm{2}$} & \multicolumn{2}{c|}{$D_\textrm{3}$} & \multicolumn{2}{c}{$D_\textrm{1}$} & \multicolumn{2}{c}{$D_\textrm{2}$} & \multicolumn{2}{c|}{$D_\textrm{3}$} \\
\hline
\multirow{2}{*}{$D_\textrm{1}$} & UNet &  67.14$\pm$&8.24 &  39.23$\pm$&8.18 &  64.77$\pm$&11.71 &   600$\pm$&232 &   1647$\pm$&215 &   938$\pm$&706 \\
& SegCLR &  69.13$\pm$&7.94 &  53.07$\pm$&8.35 &  71.13$\pm$&17.16 &   594$\pm$&245 &   1261$\pm$&224 &   827$\pm$&610 \\
\hdashline
\multirow{2}{*}{$D_\textrm{2}$} & UNet &  46.01$\pm$&6.07 &  61.30$\pm$&4.34 &  40.72$\pm$&16.08 &  1083$\pm$&480 &    961$\pm$&142 &  1445$\pm$&797 \\
    & SegCLR &  54.55$\pm$&6.27 &  64.08$\pm$&6.23 &  55.11$\pm$&22.72 &   823$\pm$&360 &    912$\pm$&158 &  1120$\pm$&826 \\
\hdashline
\multirow{2}{*}{$D_\textrm{3}$} & UNet &  43.58$\pm$&4.13 &  35.26$\pm$&7.58 &  79.45$\pm$&13.67 &   971$\pm$&279 &   2294$\pm$&311 &   652$\pm$&479 \\
& SegCLR &  40.10$\pm$&5.25 &  40.50$\pm$&9.41 &  80.21$\pm$&13.02 &  1100$\pm$&353 &  2512$\pm$&1040 &   606$\pm$&469 \\
    
\hdashline
\multirow{2}{*}{$D_\textrm{All}$} & UNet &  67.72$\pm$&7.42 &  66.25$\pm$&3.82 &  79.40$\pm$&13.25 &   580$\pm$&217 &     826$\pm$&92 &   636$\pm$&500 \\
& SegCLR & \B 69.86$\pm$&\B 7.25 &  \B 69.48$\pm$&\B 4.73 &  \B 81.39$\pm$&\B 13.18 &   \B 576$\pm$&\B 239 &    \B 818$\pm$&\B 115 &   \B 578$\pm$&\B 466 \\
\hline
\end{tabular}
}
\label{tab:results_zeroshot_all}
\end{table}

\section{Conclusion}
\label{sec:conclusion}
We have introduced SegCLR, a novel framework for learning across different domains, either labeled or unlabeled. 
Through a series of extensive experiments on OCT datasets from different clinical trials, including different acquisition devices and disease conditions, SegCLR with joint training is demonstrated to be superior to conventional training and contrastively-pretrained strategies.
SegCLR can address significant domain shifts in unsupervised domain adaptation (UDA), as demonstrated in scenarios with changing imaging devices and eye diseases, with promising results for both source and target domains.
Our experimental observations highlight the importance of training replicates in deep learning experimental work; a procedure that, while obvious, is rarely used. 
While the proposed SegCLR($P_\mathrm{a}$,$C_\mathrm{ch}$) outperforms other configurations across all the experiments, most configurations leading to similar improvements in results indicates the robustness of the proposed framework to structural changes.
Note that hyper-parameters were mostly similar across the diverse experimental settings, reaffirming the robustness.

The effectiveness of SegCLR does not rely on the availability of large amounts of unlabeled data, and it performs strikingly well even with limited or no unlabeled data from the target domain, \ie in unsupervised domain adaptation and domain generalization scenarios.
These have been demonstrated in a variety of experimental settings, with the exciting implications as follows:
First, on the contrary to what can be anticipated, very large unlabeled datasets are not required to benefit from SegCLR, and even minimal target data is helpful.
Second, impressive results can already be attained with no access to or knowledge of the target domain at all!
In such domain generalization setting, labeled data is available only from source domain(s), with additional target domains only encountered upon deployment.

The versatility of SegCLR is further demonstrated in a multi-domain training experiment where it is concurrently trained on labeled data from multiple domains. 
The results affirm that SegCLR is not only beneficial for new, unseen domains in a UDA setting, but it also yields superior results in typical supervised settings with labeled data, from both a single or multiple domains.
Accordingly, we propose SegCLR to augment conventional supervised UNet, with added generalizability and without compromise.

In summary, our findings indicate that SegCLR provides a powerful and versatile approach to tackle training and inference across multiple domains in medical imaging. 
With its capability of learning from both labeled and unlabeled data across diverse domains, SegCLR shows great promise for improving the applicability of deep learning models to complex clinical settings involving heterogeneous domains. 
Furthermore, in contrast to other UDA approaches, SegCLR can retain source domain knowledge while adapting to target domains, which can facilitate its use and adaptation for future applications in continual and incremental learning.

\bibliographystyle{IEEEtran}
\bibliography{refs}

\newpage
\onecolumn
\begin{center}
\Large {\bf{Supplementary Material\\}}
\end{center}

\setcounter{figure}{0}  
\renewcommand{\thefigure}{S\arabic{figure}}
\setcounter{table}{0}  
\renewcommand{\thetable}{S\arabic{table}}

\begin{table*}[h]
    \centering
    \caption{
    Datasets employed for the training and evaluation of models. 
    Labeled data for training is displayed as \#training+\#validation. 
    While the unlabeled data for $D_2$ and $D_3$ come from a database of numerous non-annotated volumes, the  unlabeled data for $D_1$ is considered to be the unlabeled B-scans of the labeled volumes in order to enable evaluations also on that domain.
    }
    \footnotesize
\setlength{\extrarowheight}{.3ex}
\begin{tabular}{|c|c|c|cc|cc|cc|}
\hline
\multirow{3}{*}{ \textbf{Domain} } & \multirow{3}{*}{ \textbf{Device} } & \multirow{3}{*}{ \textbf{Disease} } & \multicolumn{4}{c|}{\textbf{Training}} &
\multicolumn{2}{c|}{\textbf{Testing}} \\
\cdashline{4-9}
& & & \multicolumn{2}{c|}{\textbf{Labeled}} & \multicolumn{2}{c|}{\textbf{Unlabeled}} & \multicolumn{2}{c|}{\textbf{Labeled}} \\
\cdashline{4-9}
& & &\textbf{ B-scans} & \textbf{Volumes} & \textbf{B-scans} & \textbf{Volumes} & \textbf{B-scans} & \textbf{Volumes} \\
\hline
$D_1$ & Spectralis & nAMD & 1363+243 & 234+41 & 11\,466 & 275 & 163 & 28 \\
$D_2$ & Cirrus & nAMD & 735+125 & 122+21 & $6.8 \times 10^6$ & 53\,197 & 99 & 17 \\
$D_3$ & Spectralis & DME & 1264+226 & 228+40 & $1.1 \times 10^5$ & 4098 & 196 & 35 \\
\hline
    \end{tabular}
    \label{tab:dataset_details}
\end{table*}

\begin{table*}[h]
\centering
\caption{Datasets employed as source ($D^s$), target ($D^t$) in the experimental setting for each of the paper sections.}
\footnotesize
\begin{tabular}{|l|c|c|c|}
\hline
\multicolumn{1}{|c|}{\multirow{2}{*}{\textbf{Section \& experiment}}} & \multicolumn{2}{c|}{\textbf{Training}} & \multicolumn{1}{c|}{\multirow{2}{*}{\textbf{Evaluation}}} \\
\cline{2-3}
 & \textbf{Labeled ($D^s$)} & \textbf{Unlabeled ($D^t$)} & \\
\hline
4.1. Cross-device DA & $D_1$ & $D_2$ & $D_1$; $D_2$ \\
\hline
4.2. Cross-disease DA & $D_1$ & $D_3$ & $D_1$; $D_3$ \\
\hline
4.4. Reducing data & $D_1$ & $D_2$ & $D_1$; $D_2$ \\
\hline
4.5. Domain generalization & \makecell{$D_1$; $D_2$; $D_3$;\\ $D_1\!\cup\!D_2\!\cup\!D_3$} & N/A & \makecell{$D_1$; $D_2$; $D_3$;\\ $D_1\!\cup\!D_2\!\cup\!D_3$} \\
\hline
\end{tabular}
\label{tab:experiment_details}
\end{table*}

\begin{table*}[h]
\centering
\caption{
Absolute segmentation metrics for different tissue classes in the unsupervised domain adaptation setting with $D^s=D_1$ and $D^t=D_2$.
}
\resizebox{\textwidth}{!}{%

\begin{tabular}{|l|r@{}lr@{}lr@{}lr@{}l:r@{}lr@{}lr@{}lr@{}l|r@{}lr@{}lr@{}lr@{}l:r@{}lr@{}lr@{}lr@{}l|}
\hline
\multicolumn{1}{|c|}{\multirow{3}{*}{Model}} & \multicolumn{16}{c}{Dice [\%]} & \multicolumn{16}{|c|}{UVD [fL]} \\
\cdashline{2-33}
& \multicolumn{8}{c:}{Target domain} & \multicolumn{8}{c|}{Source domain} & \multicolumn{8}{c:}{Target domain} & \multicolumn{8}{c|}{Source domain} \\
\cdashline{2-33}
& \multicolumn{2}{c}{IRF} & \multicolumn{2}{c}{SRF} & \multicolumn{2}{c}{PED} & \multicolumn{2}{c:}{SHRM} & \multicolumn{2}{c}{IRF} & \multicolumn{2}{c}{SRF} & \multicolumn{2}{c}{PED} & \multicolumn{2}{c|}{SHRM} & \multicolumn{2}{c}{IRF} & \multicolumn{2}{c}{SRF} & \multicolumn{2}{c}{PED} & \multicolumn{2}{c:}{SHRM} & \multicolumn{2}{c}{IRF} & \multicolumn{2}{c}{SRF} & \multicolumn{2}{c}{PED} & \multicolumn{2}{c|}{SHRM}\\
\hline

UNet (UpperBound)                          &   67.25$\pm$&1.66 &   59.64$\pm$&3.31 &  60.39$\pm$&1.70 &   57.94$\pm$&3.02 &            N/A& &            N/A& &            N/A& &            N/A& &    1170$\pm$&36 &    978$\pm$&53 &    819$\pm$&46 &    877$\pm$&46 &           N/A& &         N/A& &        N/A& &        N/A& \\
UNet (Baseline)                            &   44.82$\pm$&4.90 &   44.22$\pm$&6.15 &  34.98$\pm$&7.72 &   32.91$\pm$&6.48 &  77.72$\pm$&2.07 &  63.64$\pm$&1.13 &  70.98$\pm$&0.57 &  56.20$\pm$&0.99 &    1858$\pm$&92 &  1670$\pm$&196 &  1445$\pm$&219 &   1614$\pm$&98 &     226$\pm$&11 &   815$\pm$&30 &  622$\pm$&47 &  738$\pm$&27 \\
\hdashline
CycleGAN                                   &   42.30$\pm$&4.33 &   40.93$\pm$&7.78 &  32.31$\pm$&5.48 &   31.82$\pm$&5.17 &            N/A& &            N/A& &            N/A& &            N/A& &   1813$\pm$&143 &  1786$\pm$&228 &  1498$\pm$&169 &   1643$\pm$&87 &           N/A& &         N/A& &        N/A& &        N/A& \\
SVDNA                                      &   64.36$\pm$&3.14 &   59.95$\pm$&5.33 &  53.13$\pm$&2.71 &   55.33$\pm$&1.87 &  75.73$\pm$&2.87 &  61.85$\pm$&1.81 &  70.23$\pm$&1.08 &  55.65$\pm$&1.76 &    1225$\pm$&72 &    944$\pm$&55 &    912$\pm$&64 &    972$\pm$&87 &      231$\pm$&8 &   863$\pm$&54 &  601$\pm$&29 &  743$\pm$&28 \\
\hdashline
SimCLR                                     &   47.23$\pm$&6.07 &   44.22$\pm$&7.29 &  37.56$\pm$&4.26 &   35.05$\pm$&4.13 &  76.67$\pm$&2.72 &  63.20$\pm$&1.93 &  70.74$\pm$&1.29 &  56.19$\pm$&2.40 &   1737$\pm$&165 &  1704$\pm$&165 &  1402$\pm$&124 &   1613$\pm$&98 &     231$\pm$&10 &   854$\pm$&39 &  604$\pm$&36 &  744$\pm$&35 \\
SimSiam                                    &   45.37$\pm$&3.83 &   48.94$\pm$&4.52 &  36.09$\pm$&5.45 &   34.17$\pm$&5.72 &  78.81$\pm$&1.94 &  63.26$\pm$&1.23 &  71.15$\pm$&0.77 &  56.37$\pm$&1.46 &    1800$\pm$&98 &  1630$\pm$&149 &  1395$\pm$&127 &  1566$\pm$&118 &      222$\pm$&7 &   854$\pm$&43 &  595$\pm$&44 &  742$\pm$&37 \\
\hdashline
SegCLR($P_\mathrm{a}$,$C_\mathrm{pool}$)   &   61.60$\pm$&6.64 &   60.55$\pm$&4.70 &  53.69$\pm$&2.61 &   56.54$\pm$&5.55 &  75.91$\pm$&4.55 &  63.60$\pm$&1.42 &  70.43$\pm$&1.13 &  59.05$\pm$&1.14 &   1271$\pm$&141 &  1049$\pm$&111 &  1222$\pm$&247 &  1166$\pm$&132 &      256$\pm$&6 &   887$\pm$&52 &  551$\pm$&27 &  698$\pm$&57 \\
SegCLR($P_\mathrm{s}$,$C_\mathrm{pool}$)   &   34.78$\pm$&4.71 &   35.46$\pm$&7.18 &  44.99$\pm$&9.47 &   42.32$\pm$&7.45 &  52.19$\pm$&6.97 &  47.70$\pm$&3.00 &  62.03$\pm$&2.80 &  47.03$\pm$&2.97 &   1803$\pm$&186 &  1768$\pm$&450 &  1566$\pm$&597 &  1763$\pm$&965 &     298$\pm$&13 &  1131$\pm$&54 &  961$\pm$&98 &  925$\pm$&84 \\
SegCLR($P_\mathrm{s+a}$,$C_\mathrm{pool}$) &   64.00$\pm$&6.72 &   59.16$\pm$&2.88 &  56.50$\pm$&3.98 &   58.79$\pm$&5.43 &  75.51$\pm$&3.86 &  62.79$\pm$&1.70 &  71.30$\pm$&1.37 &  57.97$\pm$&1.18 &    1249$\pm$&80 &  1067$\pm$&149 &  1099$\pm$&203 &  1237$\pm$&320 &     263$\pm$&13 &   864$\pm$&49 &  556$\pm$&28 &  706$\pm$&42 \\
SegCLR($P_\mathrm{a}$,$C_\mathrm{ch}$)     &   61.29$\pm$&7.87 &   58.30$\pm$&3.06 &  57.09$\pm$&4.06 &   56.62$\pm$&4.57 &  76.22$\pm$&5.23 &  63.20$\pm$&1.84 &  71.31$\pm$&2.11 &  61.01$\pm$&2.12 &   1307$\pm$&175 &  1076$\pm$&212 &  1096$\pm$&245 &  1170$\pm$&279 &     252$\pm$&12 &   850$\pm$&58 &  540$\pm$&23 &  696$\pm$&41 \\
SegCLR($P_\mathrm{s}$,$C_\mathrm{ch}$)     &  52.76$\pm$&10.43 &  46.87$\pm$&10.14 &  51.74$\pm$&3.14 &  51.03$\pm$&10.30 &  74.25$\pm$&2.87 &  62.46$\pm$&0.89 &  71.44$\pm$&0.82 &  57.36$\pm$&1.32 &   1476$\pm$&417 &  1502$\pm$&466 &  1148$\pm$&255 &  1174$\pm$&135 &      233$\pm$&8 &   850$\pm$&29 &  602$\pm$&31 &  766$\pm$&26 \\
SegCLR($P_\mathrm{s+a}$,$C_\mathrm{ch}$)   &   62.44$\pm$&4.73 &   59.30$\pm$&6.18 &  55.86$\pm$&2.05 &   54.78$\pm$&6.01 &  75.85$\pm$&3.82 &  63.79$\pm$&1.51 &  69.67$\pm$&1.96 &  59.86$\pm$&2.80 &   1320$\pm$&181 &  1075$\pm$&149 &  1112$\pm$&230 &  1117$\pm$&169 &     260$\pm$&17 &   881$\pm$&64 &  552$\pm$&39 &  737$\pm$&49 \\
\hline
\end{tabular}
}
\label{tab:results_avenue_harbor}
\end{table*}

\begin{table*}[!h]
\centering
\caption{
Absolute segmentation metrics for different tissue classes in the unsupervised domain adaptation setting with $D^s=D_1$ and $D^t=D_3$.
}
\resizebox{\linewidth}{!}{%
\begin{tabular}{|l|r@{}lr@{}l:r@{}lr@{}lr@{}lr@{}l|r@{}lr@{}l:r@{}lr@{}lr@{}lr@{}l|}
\hline
\multicolumn{1}{|c|}{\multirow{3}{*}{Model}} & \multicolumn{12}{c}{Dice [\%]} & \multicolumn{12}{|c|}{UVD [fL]} \\
\cdashline{2-25}
& \multicolumn{4}{c:}{Target domain} & \multicolumn{8}{c|}{Source domain} & \multicolumn{4}{c:}{Target domain} & \multicolumn{8}{c|}{Source domain} \\
\cdashline{2-25}
& \multicolumn{2}{c}{IRF} & \multicolumn{2}{c:}{SRF} & \multicolumn{2}{c}{IRF} & \multicolumn{2}{c}{SRF} & \multicolumn{2}{c}{PED} & \multicolumn{2}{c|}{SHRM} & \multicolumn{2}{c}{IRF} & \multicolumn{2}{c:}{SRF} & \multicolumn{2}{c}{IRF} & \multicolumn{2}{c}{SRF} & \multicolumn{2}{c}{PED} & \multicolumn{2}{c|}{SHRM}\\
\hline
UNet (UpperBound)                          &  65.45$\pm$&1.09 &   93.18$\pm$&0.85 &            N/A& &            N/A& &            N/A& &            N/A& &    1136$\pm$&23 &   179$\pm$&17 &           N/A& &        N/A& &        N/A& &        N/A& \\
UNet (Baseline)                            &  53.84$\pm$&3.26 &   75.71$\pm$&3.64 &  77.72$\pm$&2.07 &  63.64$\pm$&1.13 &  70.98$\pm$&0.57 &  56.20$\pm$&0.99 &    1623$\pm$&99 &   253$\pm$&24 &     226$\pm$&11 &  815$\pm$&30 &  622$\pm$&47 &  738$\pm$&27 \\
\hdashline
CycleGAN                                   &  55.08$\pm$&2.80 &   75.39$\pm$&3.90 &            N/A& &            N/A& &            N/A& &            N/A& &   1592$\pm$&110 &   247$\pm$&32 &           N/A& &        N/A& &        N/A& &        N/A& \\
SVDNA                                      &  56.40$\pm$&3.88 &   78.38$\pm$&5.28 &  77.08$\pm$&2.46 &  62.31$\pm$&1.52 &  70.56$\pm$&0.90 &  56.30$\pm$&1.08 &   1576$\pm$&127 &   227$\pm$&20 &     233$\pm$&10 &  853$\pm$&41 &  603$\pm$&30 &  744$\pm$&33 \\
\hdashline
SimCLR                                     &  53.42$\pm$&2.07 &   57.24$\pm$&7.70 &  69.80$\pm$&3.35 &  61.38$\pm$&2.30 &  68.76$\pm$&1.09 &  54.73$\pm$&1.26 &    1663$\pm$&78 &   363$\pm$&59 &     262$\pm$&11 &  852$\pm$&34 &  651$\pm$&45 &  762$\pm$&27 \\
SimSiam                                    &  54.71$\pm$&3.69 &   76.70$\pm$&5.77 &  76.46$\pm$&2.89 &  63.21$\pm$&1.18 &  71.18$\pm$&0.88 &  56.76$\pm$&1.61 &    1588$\pm$&82 &   254$\pm$&41 &     230$\pm$&12 &  831$\pm$&32 &  583$\pm$&37 &  737$\pm$&26 \\
\hdashline
SegCLR($P_\mathrm{a}$,$C_\mathrm{pool}$)   &  58.31$\pm$&5.09 &   86.68$\pm$&4.53 &  78.62$\pm$&1.93 &  63.55$\pm$&1.98 &  71.17$\pm$&1.65 &  60.60$\pm$&2.40 &   1365$\pm$&155 &   224$\pm$&54 &      253$\pm$&9 &  861$\pm$&69 &  529$\pm$&18 &  701$\pm$&36 \\
SegCLR($P_\mathrm{s}$,$C_\mathrm{pool}$)   &  50.46$\pm$&2.64 &  75.79$\pm$&12.31 &  67.74$\pm$&6.05 &  61.00$\pm$&2.09 &  70.09$\pm$&1.49 &  53.19$\pm$&1.62 &    1724$\pm$&56 &  375$\pm$&209 &     255$\pm$&11 &  854$\pm$&84 &  605$\pm$&24 &  785$\pm$&56 \\
SegCLR($P_\mathrm{s+a}$,$C_\mathrm{pool}$) &  56.45$\pm$&4.12 &   86.51$\pm$&4.18 &  78.03$\pm$&2.16 &  64.48$\pm$&2.08 &  71.06$\pm$&1.62 &  60.98$\pm$&2.91 &    1412$\pm$&85 &   279$\pm$&71 &      250$\pm$&9 &  855$\pm$&84 &  533$\pm$&46 &  694$\pm$&55 \\
SegCLR($P_\mathrm{a}$,$C_\mathrm{ch}$)     &  58.15$\pm$&3.17 &   88.89$\pm$&3.21 &  77.82$\pm$&2.38 &  63.14$\pm$&1.86 &  70.92$\pm$&2.14 &  61.11$\pm$&2.04 &    1359$\pm$&76 &   210$\pm$&54 &     251$\pm$&12 &  883$\pm$&61 &  523$\pm$&18 &  716$\pm$&47 \\
SegCLR($P_\mathrm{s}$,$C_\mathrm{ch}$)     &  56.17$\pm$&2.47 &   73.84$\pm$&8.96 &  73.78$\pm$&3.06 &  61.66$\pm$&2.36 &  70.74$\pm$&0.95 &  57.21$\pm$&2.02 &    1630$\pm$&69 &   333$\pm$&63 &      232$\pm$&8 &  881$\pm$&29 &  606$\pm$&54 &  762$\pm$&37 \\
SegCLR($P_\mathrm{s+a}$,$C_\mathrm{ch}$)   &  58.39$\pm$&3.46 &   89.60$\pm$&3.86 &  79.10$\pm$&2.48 &  63.72$\pm$&1.71 &  70.98$\pm$&1.84 &  60.88$\pm$&2.53 &    1313$\pm$&62 &   199$\pm$&68 &     258$\pm$&20 &  832$\pm$&60 &  536$\pm$&59 &  710$\pm$&66 \\
\hline
\end{tabular}
}
\label{tab:results_avenue_boulevard}
\end{table*}

\begin{table*}[h]
\centering
\caption{
Domain generalization results for models evaluated on $D_1$ after training on different $D^s$.
Absolute metrics are shown for each of the evaluated classes.
}
\resizebox{\linewidth}{!}{%
\begin{tabular}{|c|c|r@{}lr@{}lr@{}lr@{}l|r@{}lr@{}lr@{}lr@{}l|}
\hline
\multicolumn{1}{|c|}{\multirow{2}{*}{$D^s$}} & \multicolumn{1}{c|}{\multirow{2}{*}{Model}} & \multicolumn{8}{c|}{Dice [\%]} & \multicolumn{8}{c|}{UVD [fL]}  \\
\cdashline{3-18}
& & \multicolumn{2}{c}{IRF} & \multicolumn{2}{c}{SRF} & \multicolumn{2}{c}{PED} & \multicolumn{2}{c|}{SHRM} & \multicolumn{2}{c}{IRF} & \multicolumn{2}{c}{SRF} & \multicolumn{2}{c}{PED} & \multicolumn{2}{c|}{SHRM} \\
\hline
\multirow{2}{*}{$D_\textrm{1}$} & UNet &   77.72$\pm$&2.07 &  63.64$\pm$&1.13 &  70.98$\pm$&0.57 &  56.20$\pm$&0.99 &   226$\pm$&11 &    815$\pm$&30 &    622$\pm$&47 &    738$\pm$&27 \\
& SegCLR &   80.51$\pm$&1.89 &  64.25$\pm$&1.55 &  71.35$\pm$&1.15 &  60.43$\pm$&2.43 &   243$\pm$&10 &    889$\pm$&66 &    531$\pm$&21 &    714$\pm$&50 \\
    \hdashline
\multirow{2}{*}{$D_\textrm{2}$} & UNet &   43.08$\pm$&7.38 &  45.44$\pm$&4.99 &  51.56$\pm$&4.23 &  43.94$\pm$&3.66 &   424$\pm$&77 &  1566$\pm$&233 &   1021$\pm$&77 &  1319$\pm$&355 \\
& SegCLR &  57.35$\pm$&10.06 &  54.06$\pm$&4.21 &  50.36$\pm$&3.66 &  56.44$\pm$&2.38 &   354$\pm$&41 &  1054$\pm$&216 &  1038$\pm$&394 &    848$\pm$&47 \\
    
\hdashline
\multirow{2}{*}{$D_\textrm{3}$} & UNet &   46.46$\pm$&4.12 &  40.71$\pm$&0.81 &            N/&A &            N/&A &  719$\pm$&146 &   1224$\pm$&30 &          N/&A &          N/&A \\
& SegCLR &   41.58$\pm$&6.57 &  38.63$\pm$&3.21 &            N/&A &            N/&A &  857$\pm$&356 &   1342$\pm$&79 &          N/&A &          N/&A \\
\hdashline
\multirow{2}{*}{$D_\textrm{All}$} & UNet &   77.41$\pm$&1.94 &  64.76$\pm$&0.81 &  70.83$\pm$&0.84 &  57.87$\pm$&1.02 &   233$\pm$&10 &    788$\pm$&15 &    585$\pm$&33 &    715$\pm$&10 \\
& SegCLR &   79.84$\pm$&3.07 &  64.05$\pm$&2.18 &  72.42$\pm$&1.53 &  63.14$\pm$&2.31 &    235$\pm$&9 &    854$\pm$&53 &    501$\pm$&21 &    714$\pm$&53 \\
\hline
\end{tabular}
}
\label{tab:results_zeroshot_d1}
\end{table*}

\begin{table*}[h]
\centering
\caption{
Domain generalization results for models evaluated on $D_2$ after training on different $D^s$.
Absolute metrics are shown for each of the evaluated classes.
}
\resizebox{\linewidth}{!}{%
\begin{tabular}{|c|c|r@{}lr@{}lr@{}lr@{}l|r@{}lr@{}lr@{}lr@{}l|}
\hline
\multicolumn{1}{|c|}{\multirow{2}{*}{$D^s$}} & \multicolumn{1}{c|}{\multirow{2}{*}{Model}} & \multicolumn{8}{c|}{Dice [\%]} & \multicolumn{8}{c|}{UVD [fL]}  \\
\cdashline{3-18}
& & \multicolumn{2}{c}{IRF} & \multicolumn{2}{c}{SRF} & \multicolumn{2}{c}{PED} & \multicolumn{2}{c|}{SHRM} & \multicolumn{2}{c}{IRF} & \multicolumn{2}{c}{SRF} & \multicolumn{2}{c}{PED} & \multicolumn{2}{c|}{SHRM} \\
\hline
\multirow{2}{*}{$D_\textrm{1}$}  & UNet &  44.82$\pm$&4.90 &   44.22$\pm$&6.15 &  34.98$\pm$&7.72 &  32.91$\pm$&6.48 &    1858$\pm$&92 &  1670$\pm$&196 &  1445$\pm$&219 &   1614$\pm$&98 \\
& SegCLR &  59.66$\pm$&7.82 &  52.37$\pm$&10.61 &  49.93$\pm$&3.99 &  50.33$\pm$&6.67 &   1377$\pm$&154 &  1221$\pm$&292 &  1254$\pm$&227 &  1194$\pm$&186 \\
\hdashline
\multirow{2}{*}{$D_\textrm{2}$} & UNet &  67.25$\pm$&1.66 &   59.64$\pm$&3.31 &  60.39$\pm$&1.70 &  57.94$\pm$&3.02 &    1170$\pm$&36 &    978$\pm$&53 &    819$\pm$&46 &    877$\pm$&46 \\
& SegCLR &  71.10$\pm$&2.44 &   58.19$\pm$&5.28 &  66.71$\pm$&3.26 &  60.34$\pm$&2.70 &    1095$\pm$&54 &   1003$\pm$&78 &    735$\pm$&38 &    817$\pm$&77 \\
\hdashline
\multirow{2}{*}{$D_\textrm{3}$} & UNet &  41.52$\pm$&2.04 &   29.01$\pm$&5.48 &            N/&A &            N/&A &   2410$\pm$&394 &  2178$\pm$&139 &          N/&A &          N/&A \\
& SegCLR &  45.37$\pm$&6.81 &   35.62$\pm$&9.37 &            N/&A &            N/&A &  2898$\pm$&1293 &  2127$\pm$&530 &          N/&A &          N/&A \\
\hdashline
\multirow{2}{*}{$D_\textrm{All}$} & UNet &  70.81$\pm$&2.14 &   66.16$\pm$&1.84 &  66.49$\pm$&1.64 &  61.53$\pm$&2.16 &     958$\pm$&38 &    834$\pm$&33 &    738$\pm$&36 &    775$\pm$&47 \\
& SegCLR &  74.27$\pm$&1.92 &   67.80$\pm$&4.46 &  71.05$\pm$&2.72 &  64.80$\pm$&3.22 &     958$\pm$&70 &    866$\pm$&38 &    698$\pm$&69 &    752$\pm$&37 \\
\hline
\end{tabular}
}
\label{tab:results_zeroshot_d2}
\end{table*}

\begin{table*}[h]
\centering
\caption{
Domain generalization results for models evaluated on $D_3$ after training on different $D^s$.
Absolute metrics are shown for each of the evaluated classes.
Note that PED and SHRM are not relevant for $D_3$ (see Section III).
}
\resizebox{0.6\columnwidth}{!}{%
\begin{tabular}{|c|c|r@{}lr@{}l|r@{}lr@{}l|}
\hline
\multicolumn{1}{|c|}{\multirow{2}{*}{$D^s$}} & \multicolumn{1}{c|}{\multirow{2}{*}{Model}} & \multicolumn{4}{c|}{Dice [\%]} & \multicolumn{4}{c|}{UVD [fL]} \\
\cdashline{3-10}
& & \multicolumn{2}{c}{IRF} & \multicolumn{2}{c|}{SRF} & \multicolumn{2}{c}{IRF} & \multicolumn{2}{c|}{SRF} \\
\hline
\multirow{2}{*}{$D_\textrm{1}$} & UNet &  53.84$\pm$&3.26 &   75.71$\pm$&3.64 &   1623$\pm$&99 &   253$\pm$&24 \\
& SegCLR &  54.69$\pm$&2.88 &   87.58$\pm$&3.58 &   1417$\pm$&79 &   237$\pm$&66 \\
\hdashline
\multirow{2}{*}{$D_\textrm{2}$} & UNet &  29.91$\pm$&6.51 &  51.52$\pm$&15.63 &  2206$\pm$&144 &  685$\pm$&185 \\
& SegCLR &  34.81$\pm$&5.29 &  75.41$\pm$&12.11 &  1914$\pm$&108 &  326$\pm$&168 \\    
\hdashline
\multirow{2}{*}{$D_\textrm{3}$} & UNet &  66.18$\pm$&1.61 &   92.73$\pm$&0.66 &   1119$\pm$&21 &   185$\pm$&17 \\
& SegCLR &  67.76$\pm$&0.71 &   92.65$\pm$&3.62 &   1062$\pm$&44 &   150$\pm$&23 \\
\hdashline
\multirow{2}{*}{$D_\textrm{All}$} & UNet &  66.54$\pm$&1.27 &   92.27$\pm$&1.13 &   1123$\pm$&28 &   149$\pm$&14 \\
& SegCLR &  68.61$\pm$&1.00 &   94.17$\pm$&1.57 &   1032$\pm$&17 &   124$\pm$&26 \\
\hline
\end{tabular}
}
\label{tab:results_zeroshot_d3}
\end{table*}

\end{document}